\documentclass{article}

    \PassOptionsToPackage{numbers, compress}{natbib}

 \usepackage[main, final]{neurips_2026}

\author{%
  Wei Fan \\
  Virginia Tech\\
  \texttt{fanwei@vt.edu} \\
  \And
  JinYi Yoon\thanks{Corresponding authors} \\
  Inha University \\
  \texttt{jinyiyoon@inha.ac.kr} \\
  \And
  Bo Ji\footnotemark[1] \\
  Virginia Tech\\
  \texttt{boji@vt.edu} \\
}

\usepackage[utf8]{inputenc} 
\usepackage[T1]{fontenc}    
\usepackage{hyperref}       
\usepackage{url}            
\usepackage{booktabs}       
\usepackage{amsfonts}       
\usepackage{nicefrac}       
\usepackage{microtype}      
\usepackage{xcolor}         

\usepackage{algorithm}
\usepackage{algorithmic}
\usepackage{amsmath}
\usepackage{amsfonts}
\usepackage{booktabs}
\usepackage{makecell}
\usepackage{array} 
\usepackage{multirow}
\usepackage{pifont}
\usepackage{xcolor}
\usepackage{subcaption}
\usepackage{appendix}
\newcounter{challnum}
\newcounter{insightnum}
\usepackage{xspace}
\usepackage{longtable}
\usepackage{tabularx}
\usepackage{rotating}
\usepackage{pdflscape}
\usepackage{subcaption}
\usepackage{fancyhdr}
\usepackage{lastpage}
\usepackage{amssymb}
\usepackage{amsmath}
\DeclareMathOperator*{\argmax}{arg\,max}
\DeclareMathOperator*{\argmin}{arg\,min}
\usepackage{xltabular, booktabs, multirow}
\usepackage{wrapfig}
\newcommand{\mychall}{%
  \noindent\refstepcounter{challnum}\thechallnum%
}

\newcommand{\myinsight}{%
\noindent\refstepcounter{insightnum}\theinsightnum%
}

\newcommand{\name}{\text{AIM}\xspace}
\title{AIM: Adaptive Interaction in Multi-Agent Debate \\for Multimodal LLM Inference}

\begin{document}

\maketitle

\begin{abstract}
Multi-Agent Debate (MAD) improves reasoning on multimodal tasks, but its accuracy and token efficiency depend on how agents interact during debate. A largely overlooked design choice is which \emph{interaction modalities} agents use to exchange evidence. Existing methods rely on fixed text- or graph-based interaction, which fails to convey fine-grained multimodal evidence and thus degrades accuracy while inflating token cost due to verbose descriptions. To address this, we propose \textbf{A}daptive \textbf{I}nteraction in \textbf{M}ulti-Agent Debate (\name), an adaptive routing framework that dynamically selects the optimal modality combination for inter-agent interaction in multimodal MAD. Beyond text and graph, \name introduces \emph{task-dependent modality interaction}, where agents exchange targeted regions of interest from the task-specific multimodal input, such as image regions, audio segments, or spatio-temporal video clips. To determine the best combination for each instance, \name first generates a structured baseline response and extracts interpretable routing features that capture which evidence the instance requires, how confident the baseline agent is, and how much new evidence each modality would add beyond the baseline response. A lightweight router then uses these routing features to select a modality combination that avoids both \emph{under-fusion} (i.e., agents miss decisive evidence) and \emph{over-fusion} (i.e., redundant modalities inflate token cost) risks. Through extensive experiments across all eight multimodal question-answering (QA) benchmarks, spanning Visual-QA (VQA), Audio-QA, and Video-QA, \name achieves the highest accuracy. This gain is twofold: combining interaction modalities improves accuracy over the best single-modality baseline by up to 6.4\%; while the most accurate fixed combination differs across benchmarks, \name still performs the best on all benchmarks, achieving additional accuracy improvement of up to 4.2\% and reducing token cost by up to 53.4\%. Our code is available at \url{https://github.com/Fanwei100/AIM}.
\end{abstract}

\section{Introduction}
\label{sec:intro}

Multi-Agent Systems (MAS) enhance the reasoning capabilities of Large Language Model (LLM)-based agents through collaboration, critique, and iterative refinement~\cite{liang-etal-2024-encouraging,hong2023metagpt,wu2023autogen, tillmann2025literature,hsu2025group}. Recent work extends MAS to multimodal question-answering (QA) tasks, such as Visual-QA (VQA)~\cite{multiagentvqa,Hu2024MultiAgentsBO}, Audio-QA~\cite{zhifei-etal-2025-audio}, and Video-QA~\cite{kamboj2025a,kugo2025videomultiagentsmultiagentframeworkvideo}, where agents jointly reason over language and non-textual inputs (e.g., image, audio, or video). These tasks enable a wide range of real-world applications, such as medical image interpretation~\cite{lin2023medical}, speech emotion recognition in mental health monitoring~\cite{SpeechEmotion}, and autonomous driving~\cite{marcu2024lingoqa}. Among various MAS paradigms, we focus on Multi-Agent Debate (MAD), a popular approach in which agents iteratively critique and refine each other's responses~\cite{fan2025imad, li2026m3mad, adhikari2025debatingbetter}.
Despite its effectiveness, multimodal MAD incurs high token cost, as each agent must process multimodal inputs and propagate a growing history over multiple rounds~\cite{fan2025imad, li2026m3mad}. Therefore, \emph{a key challenge is how to improve the accuracy of multimodal MAD while keeping token costs low.}

A critical but largely overlooked factor driving both accuracy and token cost is the \emph{interaction modality}, the form in which agents exchange evidence during debate. Existing multimodal MAD systems rely primarily on \emph{text-based} interaction, where agents exchange natural-language rationales and critiques~\cite{ossowski2024comma}. One recent approach explores \emph{graph-based} interaction, which encodes objects and their spatial or semantic relationships using graphs~\cite{zheng2024bdog}. Despite structural differences, both modalities essentially reduce evidence to text, making them inherently lossy when representing multimodal information (e.g., an image, an audio segment, or a video clip). When a question hinges on a small region of an image, a subtle acoustic cue, or a brief temporal event in a video clip, text or graph descriptions can miss decisive details while adding verbose explanations that increase token cost. This raises a key question: \emph{how should agents select the most appropriate interaction modality for exchanging evidence?} We explore this question motivated by the following three observations.

\emph{First, task-dependent modality interaction is more effective than text or graph.} Certain task-specific evidence, such as visual details or acoustic cues, is often more faithfully conveyed in its original modality than a textual or graph-based description. Motivated by the finding that providing targeted visual regions to single reasoning agents improves VQA accuracy~\cite{Incorporating_visual, sivakumaran2025dart}, we introduce \emph{task-dependent modality interaction}, where agents directly exchange targeted Regions of Interest (RoIs) from the original multimodal input (e.g., image, audio, or video). Specifically, agents share image RoIs for VQA, audio segments for Audio-QA, and video spatio-temporal clips for Video-QA. Our preliminary experiments show that task-dependent interaction consistently outperforms both text-only and graph-only interaction across all three task families (see Insight~\ref{insight:multimodal}). 

\emph{Second, multimodal interaction provides complementary evidence that improves reasoning.} While task-dependent interaction is proven effective, it alone may not capture all the evidence needed for accurate reasoning, since each modality contributes distinct information. For example, text preserves reasoning context, graph highlights structured relations, and task-dependent interaction exposes the original multimodal evidence. We therefore consider \emph{multimodal interaction}, which selects one or more modalities from the set of text, graph, and task-dependent modalities.
We observe that combining multiple modalities further improves accuracy and can even reduce token cost by replacing long descriptions with more targeted evidence exchange and requiring fewer debate rounds to reach an answer. However, activating all modalities available is often inefficient, since it introduces redundant evidence and substantially increases token cost without consistent accuracy gain (see Insight~\ref{insight:combination}).

\emph{Third, the best modality combination varies across instances, motivating adaptive selection.} 
A fixed modality combination is often insufficient as different questions demand different types of evidence. The best combination depends on the type of question, the demands of reasoning, and the available multimodal cues. These observations motivate \emph{adaptive modality selection} for each instance (see Insight~\ref{insight:adaptive_selection}). Specifically, adaptive modality selection routes each instance to one modality combination before the debate begins. However, realizing such \emph{modality routing} introduces two key challenges.

\noindent(\textbf{C\mychall\label{chall:first}}) \emph{How to efficiently extract informative routing signals from the initial response of a baseline agent?} 
These signals must capture the question's evidence requirements, the baseline answer's uncertainty, and the value of each modality, without repeated sampling or extra LLM queries.

\noindent(\textbf{C\mychall\label{chall:second}}) \emph{How to design a routing mechanism that identifies the optimal modality combination for accurate and efficient inference?} The mechanism must guard against two complementary risks: \emph{(i) under-fusion}: the selected combination misses decisive evidence and induces hallucination; \emph{(ii) over-fusion}: extra modalities add little gain but increase token cost without proportional accuracy improvement.

To address these challenges, we propose \textbf{\underline{A}}daptive \textbf{\underline{I}}nteraction in \textbf{\underline{M}}ulti-Agent Debate (\textbf{\name}), which, to the best of our knowledge, is the first adaptive routing framework that dynamically selects the interaction modality in multimodal MAD. We summarize our main contributions as follows:

\emph{First}, we present a systematic study of effective inter-agent interaction in multimodal MAD and identify interaction modality as a key design choice. We introduce \emph{task-dependent modality interaction} as an effective modality beyond text and graphs and motivate \emph{adaptive modality selection} by the instance-dependent nature of the optimal modality combination.

\emph{Second}, we propose \name to address (\textbf{C\ref{chall:first}}) and (\textbf{C\ref{chall:second}}) (see Fig.~\ref{fig:amid_overview}). For (\textbf{C\ref{chall:first}}), we design a \emph{structured modality-routing prompt} (Phase~{\large \textcircled{\small 1}}) that elicits routing signals from a single baseline agent, and extract \emph{interpretable routing features} (Phase~{\large \textcircled{\small 2}}) that characterize under-fusion and over-fusion risks. For (\textbf{C\ref{chall:second}}), we design an \emph{adaptive modality router} (Phase~{\large \textcircled{\small 3}}) that selects the modality combination with high accuracy at the lowest token cost per instance. For Audio-QA and Video-QA, we also introduce \emph{temporal compression} to further reduce token cost and improve accuracy.

\emph{Third}, our evaluation results show that \name achieves the highest accuracy on all eight benchmarks spanning VQA, Audio-QA, and Video-QA. This gain is twofold: combining interaction modalities improves accuracy over the best single-modality baseline by up to 6.4\%; while the most accurate fixed combination differs across benchmarks ($\mathrm{T+X}$ on four benchmarks, $\mathrm{T+G+X}$ on three, and $\mathrm{G+X}$ on one), \name still performs the best on all benchmarks, achieving additional accuracy improvement of up to 4.2\% and reducing token cost by up to 53.4\%.
\section{Related Work}
\label{sec:related_work}

We categorize the most relevant literature into two groups and then position our work relative to both.

\paragraph{Multimodal MAD.}

Multi-agent reasoning has been extended from text-only to multimodal tasks, including VQA~\cite{multiagentvqa,Hu2024MultiAgentsBO}, Audio-QA~\cite{zhifei-etal-2025-audio}, Video-QA~\cite{kamboj2025a,kugo2025videomultiagentsmultiagentframeworkvideo}, document understanding~\cite{han2025mdocagent,lassoued2026orca}, medical multimodal VQA~\cite{thakrar2025architectingclinicalcollaborationmultiagent,xia2026mmedagentrl}, and multimodal misinformation detection~\cite{lakara2025madsherlock,yang2025ramaretrievalaugmentedmultiagentframework}. Within this broader paradigm, MAD has emerged as a structured reasoning scheme in which agents iteratively critique and refine each other's responses, often improving multimodal reasoning quality~\cite{liang-etal-2024-encouraging,chan2023chateval,li2026m3mad}. Existing multimodal MAD studies fall into two directions. \emph{Accuracy-driven} methods enrich the debate process to improve answer quality. For example, \textsc{BDoG} introduces graph-based interaction alongside text~\cite{zheng2024bdog}, \textsc{DART} recruits specialized visual tools based on agent disagreement~\cite{sivakumaran2025dart}, and \textsc{WISE} employs heterogeneous expert debaters~\cite{cherian2025wise}. \emph{Efficiency-driven} methods aim to reduce token cost. For instance, \textsc{Cross-Modal Memory Compression} converts long debate histories into compact visual representations~\cite{wu2026crossmodal}, iMAD selectively triggers debate only when it is beneficial~\cite{fan2025imad}, and \textsc{ConSensus} replaces iterative debate with single-round interaction~\cite{yoon2026consensus}.

\paragraph{Modalities for Agent Interaction.}
The effectiveness of multi-agent collaboration depends heavily on the interaction modality through which agents exchange information. Most multimodal systems rely on natural-language transcripts. \textsc{COMMA} benchmarks this text-centric paradigm, showing that text-only interaction is prevalent but struggles to preserve fine-grained multimodal evidence in complex multimodal tasks~\cite{ossowski2024comma}. To address this, recent work explores alternative interaction modalities.
\textsc{BDoG} introduces blueprint graphs for structured spatial and relational grounding in VQA~\cite{zheng2024bdog}, and \textsc{SpeechAgents} shows that native speech and audio signals can function directly as inter-agent messages in audio-centric domains~\cite{zhang2024speechagents}. Beyond text, graph, and audio interaction, a separate line of work explores latent-space interaction. Methods such as \textsc{Interlat}~\cite{du2026enabling} and \textsc{LatentMAS}~\cite{zou2025latent} directly pass hidden representations between agents to reduce information loss. However, latent interaction requires access to the underlying model's hidden states and weights, making it incompatible with black-box APIs (e.g., GPT, Gemini, and Claude). Although text, graph, and audio interactions have been considered, it remains largely unexplored how to effectively exchange original multimodal evidence between agents. Moreover, existing methods typically use a fixed interaction modality in advance rather than dynamically selecting the modality for each instance.

\paragraph{Position of Our Work.}
Our work addresses a central design problem in multimodal MAD: \emph{How to design inter-agent interaction so that the system achieves high accuracy while remaining token-efficient?} Our work is orthogonal to prior accuracy-driven and efficiency-driven multimodal MAD methods. One can retain these agent designs, debate schedules, and history-compression strategies while adopting our \emph{adaptive modality interaction}. Unlike methods with fixed interaction designs or white-box latent interaction, our framework operates entirely in the prompt space and is compatible with both white-box and black-box LLM agents. We aim to dynamically select the interaction modality for each instance to improve accuracy while remaining token-efficient.

\section{Key Insights}
\label{sec:insight}

The interaction modality affects both accuracy and token cost in multimodal MAD. In this section, we present three key insights that motivate the design of \name: (i) task-dependent modality interaction improves accuracy; (ii) combining modalities further improves both accuracy and token efficiency; (iii) the best 
modality combination varies across instances, motivating adaptive modality selection.

\begin{table*}[t]
\centering
\caption{Accuracy (\%) and average token cost of all seven modality combinations across VQA (GQA), Audio-QA (DCASE), and Video-QA (NextQA) tasks, where $\mathrm{T}$, $\mathrm{G}$, and $\mathrm{X}$ denote text-based, graph-based, and task-dependent modality (e.g., image, audio, or video) interactions, respectively.
}
\label{tab:all-modality-combinations}
\resizebox{\textwidth}{!}{%
\setlength{\tabcolsep}{3pt}
\begin{tabular}{l*{7}{cc}}
\toprule
\multirow{4}{*}{\textbf{Task}}
& \multicolumn{6}{c}{\textbf{One Modality}}
& \multicolumn{6}{c}{\textbf{Two Modalities}}
& \multicolumn{2}{c}{\textbf{All}} \\
\cmidrule(lr){2-7} \cmidrule(lr){8-13} \cmidrule(lr){14-15}
& \multicolumn{2}{c}{$\mathrm{T}$}
& \multicolumn{2}{c}{$\mathrm{G}$}
& \multicolumn{2}{c}{$\mathrm{X}$}
& \multicolumn{2}{c}{$\mathrm{T+G}$}
& \multicolumn{2}{c}{$\mathrm{T+X}$}
& \multicolumn{2}{c}{$\mathrm{G+X}$}
& \multicolumn{2}{c}{$\mathrm{T+G+X}$} \\
\cmidrule(lr){2-3}
\cmidrule(lr){4-5}
\cmidrule(lr){6-7}
\cmidrule(lr){8-9}
\cmidrule(lr){10-11}
\cmidrule(lr){12-13}
\cmidrule(lr){14-15}
& \textbf{Acc} & \textbf{Tkn}
& \textbf{Acc} & \textbf{Tkn}
& \textbf{Acc} & \textbf{Tkn}
& \textbf{Acc} & \textbf{Tkn}
& \textbf{Acc} & \textbf{Tkn}
& \textbf{Acc} & \textbf{Tkn}
& \textbf{Acc} & \textbf{Tkn} \\
\midrule
VQA
& 68.9 & 6,615
& 60.6 & 8,921
& 71.2 & 6,455
& 70.8 & 6,997
& \textbf{77.6} & \textbf{6,060}
& 76.4 & 6,612
& 77.5 & 7,833 \\

Audio-QA 
& 47.6 & \textbf{3,772}
& 44.4 & 7,145
& 48.7 & 4,137
& 48.0 & 4,960
& \textbf{50.9} & 3,967
& 48.9 & 4,266
& 49.8 & 5,196 \\

Video-QA
& 72.4 & \textbf{39,814}
& 74.2 & 48,647
& 75.2 & 100,871
& 74.4 & 43,901
& 76.4 & 90,193
& 74.4 & 97,112
& \textbf{77.2} & 99,765 \\

\bottomrule
\end{tabular}%
}
\end{table*}

\paragraph{Insight \protect\myinsight\label{insight:multimodal}: Task-dependent modality interaction is more effective than text or graph for multimodal MAD.} 
In many multimodal reasoning tasks, decisive evidence resides in a specific part of the original input, such as an image region in VQA, an audio segment in Audio-QA, or a spatio-temporal clip in Video-QA. We propose task-dependent modality interaction, where agents exchange targeted RoIs from the original multimodal input (e.g., image, audio, or video) during debate. As shown in Table~\ref{tab:all-modality-combinations}, task-dependent interaction ($\mathrm{X}$) consistently achieves higher accuracy than text-based ($\mathrm{T}$) and graph-based ($\mathrm{G}$) interaction across all three tasks, revealing that \emph{the original multimodal evidence improves reasoning}. As a VQA sanity check, sharing only raw bounding-box coordinates reduces tokens but causes a substantial accuracy drop, since coordinates do not reveal the visual content inside the region~\cite{Incorporating_visual}. Thus, the accuracy gain comes from exposing the actual multimodal evidence, not merely pointing to its location. However, the task-dependent modality alone may still miss the evidence that text or graph carries, raising the question of whether combining modalities helps.

\paragraph{Insight \protect\myinsight\label{insight:combination}: Combining modalities provides complementary evidence, but using all modalities is often wasteful.} Each modality contributes unique information: text preserves reasoning context, graph highlights structural relationships, and task-dependent modality exposes the original visual, acoustic, or spatio-temporal evidence. Multimodal interaction therefore provides more comprehensive evidence than any single modality. As shown in Table~\ref{tab:all-modality-combinations}, $\mathrm{T+X}$ improves over $\mathrm{X}$ on GQA, DCASE, and NextQA in both accuracy and token cost. On GQA, accuracy rises from 71.2\% to 77.6\% while token cost drops from 6,455 to 6,060. Hence, \emph{combining modalities improves both accuracy and token efficiency beyond the task-dependent modality alone.}
While multimodal interaction is beneficial, using all modalities at all times may be inefficient. In Table~\ref{tab:all-modality-combinations}, $\mathrm{T+X}$ achieves slightly higher accuracy than $\mathrm{T+G+X}$ (77.6\% vs.\ 77.5\%) and uses 22.6\% fewer tokens on GQA. On DCASE, $\mathrm{T+X}$ achieves higher accuracy than $\mathrm{T+G+X}$ (50.9\% vs. 49.8\%) at a lower token cost. 
Therefore, \emph{using all modalities may incur high costs without improving accuracy}. 
This motivates the need for principled modality-combination selection.

\paragraph{Insight \protect\myinsight\label{insight:adaptive_selection}: The best modality combination varies across instances, motivating adaptive selection.}
\begin{wrapfigure}{r}{0.5\textwidth}
    \centering
    \vspace{-2.5em}
    \includegraphics[width=0.5\textwidth]{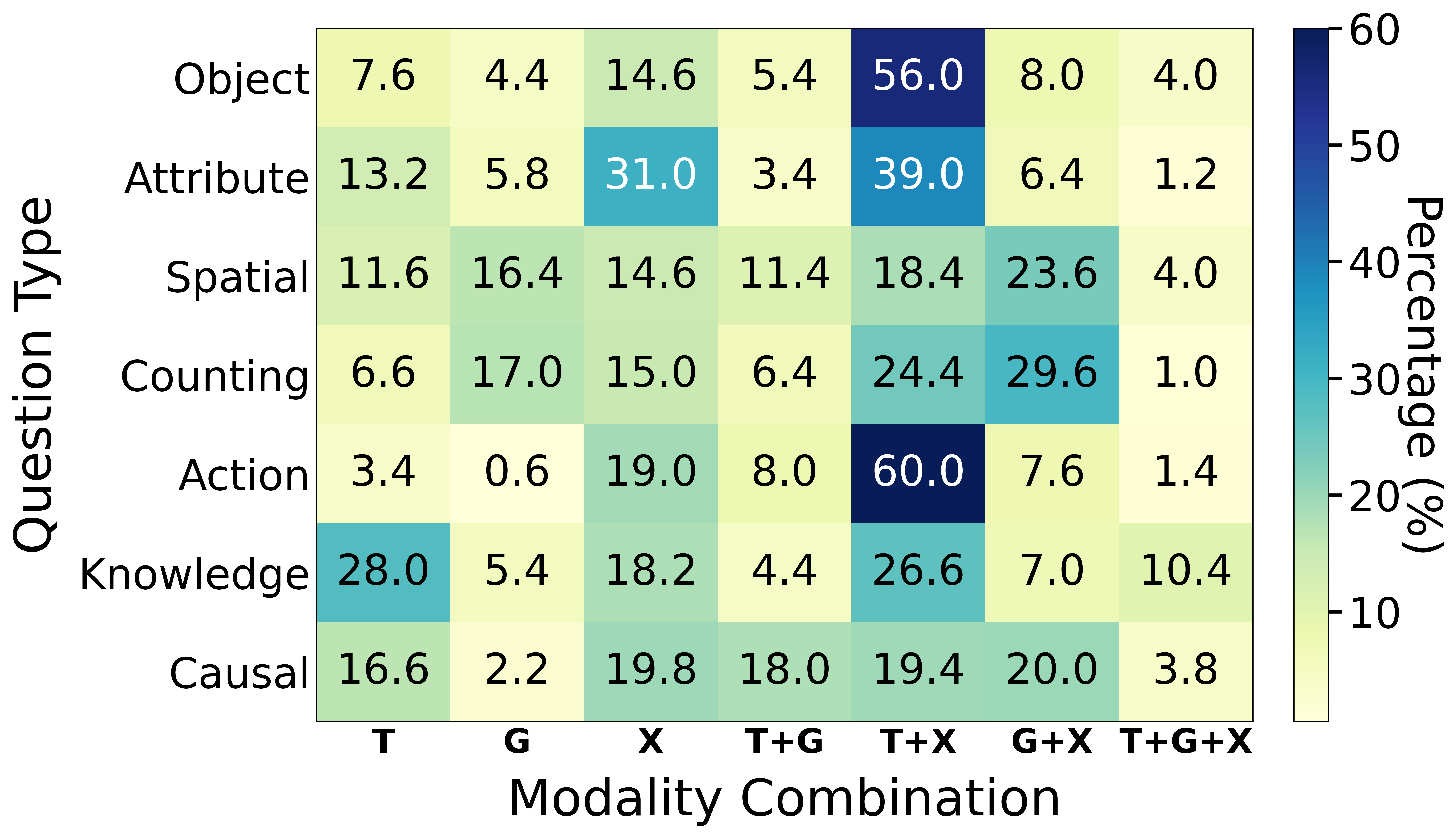}
    \caption{Best modality combination distribution across seven VQA question types.}
    \vspace{-0.1em}
    \label{fig:modality_heatmap}
\end{wrapfigure}
Although modality-combination selection is essential, no single fixed combination performs best across all instances.
To illustrate this, we define the best combination for each instance as the correct one with the lowest token cost, or the lowest-cost one if none is correct. 
As shown in Fig.~\ref{fig:modality_heatmap} for seven VQA question types,
the best combination distribution is highly non-uniform: $\mathrm{T+X}$ dominates object, attribute, and action questions; $\mathrm{G+X}$ dominates spatial and counting questions; knowledge and causal questions spread across multiple combinations. We observe similar instance-dependent patterns in Audio-QA and Video-QA (see Appendix~\ref{sec:oracle-distribution-appendix}).
These observations suggest selecting the modality combination for each instance based on its evidence requirements.

\begin{figure*}[t]
    \centering
    \includegraphics[width=\textwidth]{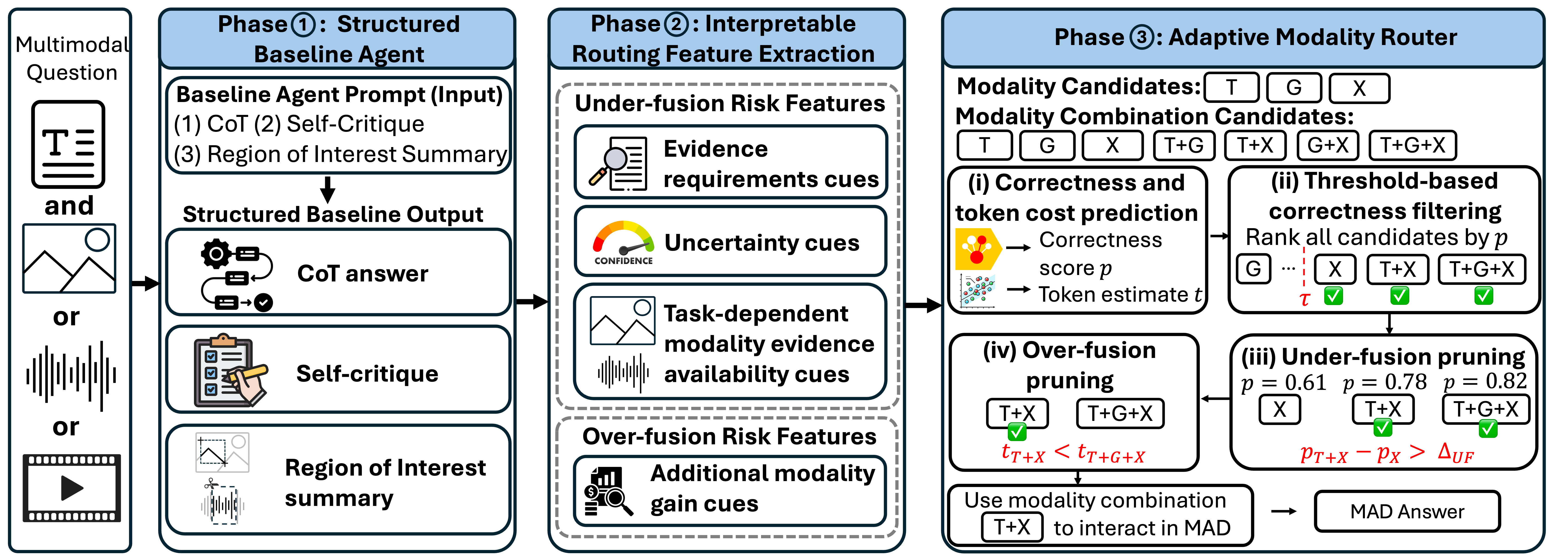}
    \caption{Overview of \name.}
    \label{fig:amid_overview}
\end{figure*}

\section{Our Design: \name}
\label{sec:method}
In this section, we present \name, an adaptive routing framework that dynamically selects the best modality combination for inter-agent interaction at the instance level. \name targets three design goals: \emph{accuracy improvement}, \emph{token efficiency}, and \emph{instance-level adaptivity}. To achieve these goals, the modality selection must guard against two risks. The first is \emph{under-fusion}, where the selected modality combination misses decisive evidence, forcing agents to bridge gaps with unsupported inference that increases hallucination and degrades accuracy. We verify this using FaithScore~\cite{jing-etal-2024-faithscore}, a widely used hallucination metric (see Section~\ref{sec:faithscore}). As shown in Table~\ref{tab:faithscore-underfusion}, single-modality settings yield much lower FaithScores (indicating more hallucination) than richer combinations, suggesting that insufficient modality evidence could be a major source of hallucination in multimodal MAD. The second risk is \emph{over-fusion}, where additional modalities introduce little new evidence while lengthening prompts and debate histories. This would increase token costs without yielding commensurate improvements in accuracy.
In Table~\ref{tab:all-modality-combinations}, the full-modality combination often costs more tokens than other combinations without consistently improving accuracy. To avoid both risks, we need to address two challenges: (i) extracting informative routing signals from a single-agent response and (ii) selecting a modality combination that provides sufficient evidence while using necessary modalities only.

As illustrated in Fig.~\ref{fig:amid_overview}, \name addresses these challenges through three key designs: 
Phase~{\large \textcircled{\small 1}}: \emph{structured baseline agent},
which produces an initial answer with compact uncertainty and per-modality evidence cues, informing downstream routing (Section~\ref{sec:baseline agent});
Phase~{\large \textcircled{\small 2}}: \emph{interpretable routing feature extraction}, which extracts routing signals from the baseline response to characterize under-fusion and over-fusion risks (Section~\ref{sec:features});
Phase~{\large \textcircled{\small 3}}: \emph{adaptive modality router}, which selects the best modality combination and executes the corresponding MAD process (Section~\ref{sec:router}). For Audio-QA and Video-QA, the task-dependent modality has an explicit temporal dimension. Hence, we introduce \emph{temporal compression} to further reduce token cost and potentially improve accuracy by compressing routed audio segments and video region clips during debate (Section~\ref{sec:temporal-compression}).

\subsection{Phase~{\large \textcircled{\small 1}}: Structured Baseline Agent}
\label{sec:baseline agent}

Given an input question $q$ with its multimodal input, \name prompts a single baseline agent to return a structured JSON response with three components. (\romannumeral1) An \emph{initial Chain-of-Thought (CoT) answer with a self-reported confidence score}, where the CoT provides the answer with reasoning and the confidence score quantifies its uncertainty. (\romannumeral2) A \emph{self-critique with a missing-evidence statement}, stating the strongest reason why the answer could be wrong and what additional information would confirm or refute it. This captures residual uncertainty and the evidence gap that the baseline agent cannot resolve. (\romannumeral3) Up to $K$ \emph{task-dependent modality summaries}, each describing a candidate region with coordinates normalized to $[0,1]$ for VQA, a segment with start and end timestamps for Audio-QA, or a spatio-temporal clip with both a spatial bounding box and timestamps for Video-QA. Each region is paired with a brief caption and a confidence score. These components expose routing signals for detecting under-fusion (i.e., missing evidence) and over-fusion (i.e., modality redundancy).

\subsection{Phase~{\large \textcircled{\small 2}}: Interpretable Routing Feature Extraction}
\label{sec:features}

From the structured baseline output, we extract interpretable features in four groups aligned with the two risks. The first three groups detect \emph{under-fusion} risk: (\romannumeral1) \emph{evidence requirement cues} include the question type, referenced entities, and missing-evidence statement, which indicate what kind of evidence the question needs, such as spatial/OCR cues for VQA, speech/event cues for Audio-QA, and action/motion cues for Video-QA; (\romannumeral2) \emph{uncertainty cues} estimate hallucination risk from the confidence score, self-critique, hedge terms, and whether the CoT grounds its answer in instance-specific evidence rather than generic priors; (\romannumeral3) \emph{task-dependent modality evidence availability cues} estimate whether the missing evidence is likely present in the task-dependent input, using region-level signals such as relevance, confidence, entity coverage, and spatial or temporal distribution. The fourth group reduces \emph{over-fusion} risk: (\romannumeral4) \emph{additional modality gain cues} estimate how much useful new evidence each modality would add to the baseline, using signals such as coverage of key entities, relationships, events, or temporal context, overlap with the current baseline reasoning, and whether the evidence is question-relevant. We provide the detailed feature list in Appendix~\ref{sec:shap-analysis}.

\subsection{Phase~{\large \textcircled{\small 3}}: Adaptive Modality Router}  
\label{sec:router}
For each question $q$, the router takes the routing feature vector $\mathbf{z}_q\in\mathbb{R}^{d}$ extracted in Phase~{\large \textcircled{\small 2}}, where $d$ is the feature dimension, and makes a routing decision $a^{*}(q)$ over the candidate set of seven options:
\[
\mathcal{A}:=\{\mathrm{T},\mathrm{G},\mathrm{X},\mathrm{T+G},\mathrm{T+X},\mathrm{G+X},\mathrm{T+G+X}\},
\]
where $\mathrm{T}$, $\mathrm{G}$, and $\mathrm{X}$ denote \textbf{T}ext-based, \textbf{G}raph-based, and task-dependent modality-based interaction, respectively. Depending on the task, $\mathrm{X}$ is image RoIs for VQA, audio segments for Audio-QA, and spatio-temporal clips for Video-QA. A combined option lets agents interact through more than one modality. The router selects one of the seven options in $\mathcal{A}$ rather than deciding on any single modality separately, since the value of one modality depends on which others are used together. For example, two modalities can jointly give more concise evidence with fewer tokens than either alone, while their interaction can improve or hurt accuracy, depending on how the information is combined. To avoid both under-fusion and over-fusion, the router proceeds with four sequential steps as follows.

(\romannumeral1) \emph{Correctness and token-cost prediction}. For each question $q$ and each candidate $a\in\mathcal{A}$, the trained router uses $\mathbf{z}_q$ to predict two quantities: a calibrated correctness probability $p(q,a)\in[0,1]$ from a binary correctness classifier and an instance-specific token-cost estimate $c(q,a)\in\mathbb{R}^{+}$ from a regression model, both trained offline on auxiliary datasets. These two estimates capture the likelihood of producing a correct answer and the expected token cost for each candidate.

(\romannumeral2) \emph{Threshold-based correctness filtering}. The router only retains candidates whose estimated correctness exceeds a tuned threshold $\tau\in[0,1]$, denoted by
\[
\mathcal{C}(q):=\{a\in\mathcal{A}:p(q,a)\ge\tau\}.
\]
This removes clearly implausible candidates so that subsequent pruning and cost-based comparison operate on a reduced and more competitive candidate set. If $\mathcal{C}(q)=\emptyset$, meaning that no candidate reaches the correctness threshold, the router falls back to the following:
\[
a^{\mathrm{fb}}(q):=\argmax_{a\in\mathcal{A}}p(q,a),
\]
i.e., the modality combination with the highest predicted correctness and thus the most likely to answer correctly among the available options. In this fallback case, the router skips the subsequent steps and directly takes the fallback combination as the final modality routing decision, i.e., $a^{*}(q)=a^{\mathrm{fb}}(q)$.

(\romannumeral3) \emph{Under-fusion pruning}. Threshold-based filtering ensures that every candidate in $\mathcal{C}(q)$ is individually plausible, but it does not compare structurally related combinations. In particular, a smaller combination may pass filtering even though adding modalities yields a much higher correctness probability. To correct this, the router compares nested combinations. Let $a \subset s$ denote that $s$ contains all modalities in $a$ and at least one additional modality. The router removes a smaller candidate $a \in \mathcal{C}(q)$ if there exists a larger combination $s \in \mathcal{C}(q)$ with $a \subset s$ such that
\[
p(q,s) - p(q,a) > \Delta_{\mathrm{UF}},
\]
where $\Delta_{\mathrm{UF}} > 0$
is a tuned under-fusion (UF) margin. A large correctness gap indicates that the smaller combination is likely missing decisive evidence, so it should not remain competitive merely because it uses fewer tokens. The resulting candidate set, denoted by $\mathcal{C}'(q)$, is defined as follows:

\[
\mathcal{C}'(q) := \bigl\{ a \in \mathcal{C}(q) : \nexists\, s \in \mathcal{C}(q) ~\text{such that}~ a \subset s ~\text{and}~ p(q,s) - p(q,a) > \Delta_{\mathrm{UF}} \bigr\}.
\]

(\romannumeral4) \emph{Over-fusion pruning via token-cost-based final selection}. After under-fusion pruning, the remaining candidates in $\mathcal{C}'(q)$ have already passed the correctness screen and are no longer obviously missing critical evidence. The router then prunes over-fused candidates and makes the final decision in a single action, by selecting the candidate with the lowest predicted token cost:
\[
a^{*}(q) := \argmin_{a \in \mathcal{C}'(q)} c(q,a).
\]
If multiple candidates share the same lowest predicted token cost, the router selects the one with the highest predicted correctness probability $p(q,a)$. This step naturally discourages over-fusion: selecting the lowest-cost candidate among the remaining plausible ones prevents the router from including extra modalities that increase token cost without significant accuracy gain. The selected combination $a^{*}(q)$ is then used to configure the modality interaction for this specific instance $q$ in MAD. The corresponding debate process is executed following the standard MAD procedure.

\subsection{Temporal Compression for Task-Dependent Interaction Involving Audio or Video}
\label{sec:temporal-compression}

For Audio-QA and Video-QA, where $\mathrm{X}$ has an explicit temporal dimension, \name further decides \emph{how much} to compress the routed evidence.
Since LLM agents tokenize audio and video at fixed rates, compressing routed audio segments or video spatio-temporal clips by a factor $r>1$ reduces token cost beyond adaptive routing (see Appendix~\ref{sec:impl-details}). 
Beyond token savings, an appropriate $r$ can also improve accuracy by removing redundant temporal context while preserving core semantic content, helping agents focus on critical evidence. We select $r$ based on the temporal density of the input to avoid under-compression (which wastes tokens) and over-compression (which makes inputs too dense to process and degrades accuracy).

\section{Evaluation}
\label{sec:evaluation}

We evaluate \name on eight multimodal QA benchmarks across VQA, Audio-QA, and Video-QA. In the main paper, we compare \name with all fixed-modality baselines (Section~\ref{sec:main-results}), analyze its hallucination severity via FaithScore (Section~\ref{sec:faithscore}), and compare \name with four single-group adaptive routers (Section~\ref{sec:routing-analysis}). In the appendix, we further provide the feature importance analysis (Appendix~\ref{sec:shap-analysis}), best combination analysis (Appendix~\ref{sec:oracle-distribution-appendix}), latency and cost analyses (Appendix~\ref{sec:cost-analyses}), ablation studies (Appendix~\ref{sec:ablation-studies}), cross-LLM-backbone results (Appendix~\ref{sec:gpt-backbone}), and robustness analysis (Appendix~\ref{sec:robustness}).

\paragraph{Datasets.} We train a separate modality router per task family on auxiliary datasets disjoint from all evaluation benchmarks: (i) Visual7W~\cite{visual7W} and CLEVR~\cite{CLEVR} for VQA; (ii) AudioMCQ~\cite{he2025measuring} for Audio-QA; (iii) Video-MME~\cite{fu2025video} for Video-QA. We use 500 auxiliary instances for VQA and 250 each for Audio-QA and Video-QA. We choose these datasets as they collectively cover diverse question types and modality requirements within each task family.

We evaluate on eight benchmarks: GQA~\cite{gqa}, OKVQA~\cite{marino2019okvqa}, VCR~\cite{zellers2019vcr}, and VQA-v2~\cite{balanced_vqa_v2} for VQA; Clotho~\cite{lipping2022clotho} and DCASE~\cite{heaudio} for Audio-QA; IntentQA~\cite{li2023intentqa} and NextQA~\cite{xiao2021next} for Video-QA. We do not use these datasets for router training or threshold tuning (see Appendix~\ref{sec:impl-details} for dataset details).

\paragraph{Baselines.} We compare \name against all seven fixed modality interaction configurations. For single-modality interaction, we include: (\romannumeral1) MAD~($\mathrm{T}$)~\cite{liang-etal-2024-encouraging}: standard text-only MAD; (\romannumeral2) Graph~($\mathrm{G}$)~\cite{zheng2024bdog}: blueprint graph-based interaction; (\romannumeral3) Task-dependent~($\mathrm{X}$): instantiated as image RoIs for VQA, audio segments for Audio-QA, and spatio-temporal video clips for Video-QA. For two-modality and full-modality interaction, we include: (\romannumeral4) $\mathrm{T+G}$~\cite{zheng2024bdog}, (\romannumeral5) $\mathrm{T+X}$, (\romannumeral6) $\mathrm{G+X}$, and (\romannumeral7) $\mathrm{T+G+X}$.

\paragraph{Metrics.}
We evaluate each method using three metrics:
(\romannumeral1) answer accuracy (\textbf{Acc}), the correctness of the final answer after debate;
(\romannumeral2) average token cost (\textbf{Tkn}), the total input and output tokens per instance, including both the structured baseline response and the subsequent MAD process
(Appendix~\ref{sec:token-breakdown}); (\romannumeral3) FaithScore~\cite{jing-etal-2024-faithscore}, assessing hallucination level, where a higher score indicates better-grounded answers. We also report the latency and training cost of \name (see Appendix~\ref{sec:cost-analyses}).

\paragraph{Implementation Details.}
For LLM-based prompting, we use Gemini 2.5 Flash for the primary results and use GPT-5.4-nano for cross-model evaluations (see Appendix~\ref{sec:gpt-backbone} for details). 
We provide the full MAD protocol, LLM agent configuration, graph interaction details, dataset details, router configurations, feature extraction details, and temporal compression procedures in Appendix~\ref{sec:impl-details}.

\begin{table*}[t]
\centering
\caption{Accuracy ({Acc}) and average token cost ({Tkn}) across eight multimodal QA benchmarks. Accuracy is measured in percentage (\%), and tokens denote the average token cost per instance, where K denotes thousands. The highest Acc and the lowest Tkn per dataset are in bold.}
\label{tab:main-results}

\resizebox{\textwidth}{!}
{%
\setlength{\tabcolsep}{2pt}
\begin{tabular}{lccccccccccccccccccccc}
\toprule
\multicolumn{1}{c}{\multirow{4}{*}{\textbf{Dataset}}}
& \multicolumn{6}{c}{\textbf{One Modality}}
& \multicolumn{6}{c}{\textbf{Two Modalities}}
& \multicolumn{2}{c}{\textbf{All}}
& \multicolumn{2}{c}{\textbf{Adaptive}} \\
\cmidrule(lr){2-7} \cmidrule(lr){8-13} \cmidrule(lr){14-15} \cmidrule(lr){16-17}
& \multicolumn{2}{c}{$\mathrm{T}$}
& \multicolumn{2}{c}{$\mathrm{G}$}
& \multicolumn{2}{c}{$\mathrm{X}$}
& \multicolumn{2}{c}{$\mathrm{T+G}$}
& \multicolumn{2}{c}{$\mathrm{T+X}$}
& \multicolumn{2}{c}{$\mathrm{G+X}$}
& \multicolumn{2}{c}{$\mathrm{T+G+X}$}
& \multicolumn{2}{c}{\name} \\
\cmidrule(lr){2-3}
\cmidrule(lr){4-5}
\cmidrule(lr){6-7}
\cmidrule(lr){8-9}
\cmidrule(lr){10-11}
\cmidrule(lr){12-13}
\cmidrule(lr){14-15}
\cmidrule(lr){16-17}
& \textbf{Acc} & \textbf{Tkn}
& \textbf{Acc} & \textbf{Tkn}
& \textbf{Acc} & \textbf{Tkn}
& \textbf{Acc} & \textbf{Tkn}
& \textbf{Acc} & \textbf{Tkn}
& \textbf{Acc} & \textbf{Tkn}
& \textbf{Acc} & \textbf{Tkn}
& \textbf{Acc} & \textbf{Tkn} \\
\midrule

GQA
& 68.9 &6,615
& 60.6 &8,921
& 71.2 &6,455
& 70.8 &6,997
& 77.6 &\textbf{6,060}
& 76.4 &6,612
& 77.5 &7,833
& \textbf{80.1} &6,334 \\

OKVQA
& 89.8 &7,803
& 89.6 &9,871
& 90.4 &7,782
& 89.8 &8,114
& 93.4 &\textbf{5,928}
& 91.2 &7,791
& 93.5 &9,773
& \textbf{93.8} &6,775 \\

VCR
& 77.6 &12,572
& 78.4 &12,588
& 79.2 &12,202
& 77.6 &10,062
& 79.8 &7,869
& 81.1 &10,861
& 79.6 &10,977
& \textbf{84.2} &\textbf{7,788} \\

VQA-v2
& 81.0 &8,796
& 77.6 &10,583
& 81.7 &7,530
& 81.0 &9,332
& 84.0 &7,012
& 81.7 &8,185
& 81.1 &9,317
& \textbf{87.7} &\textbf{7,006} \\
\midrule

Clotho
& 39.5 &\textbf{3,692}
& 41.1 &4,103
& 43.1 &4,268
& 40.2 &4,237
& 43.7 &3,897
& 44.2 &3,774
& 45.1 &4,416
& \textbf{46.8} &3,795 \\

DCASE
& 47.6 &\textbf{3,772}
& 44.4 &7,145
& 48.7 &4,137
& 48.0 &4,960
& 50.9 &3,967
& 48.9 &4,266
& 49.8 &5,196
& \textbf{51.8} &3,891 \\
\midrule

IntentQA
& 64.0 &\textbf{44.3K}
& 61.1 &66.2K
& 68.4 &109.7K
& 66.8 &56.2K
& 71.2 &98.2K
& 70.0 &102.4K
& 70.8 &100.2K
& \textbf{75.4} &45.8K \\

NextQA
& 72.4 &\textbf{39.8K}
& 74.2 &48.6K
& 75.2 &100.9K
& 74.4 &43.9K
& 76.4 &90.2K
& 74.4 &97.1K
& 77.2 &99.8K
& \textbf{78.8} &48.6K \\
\bottomrule

\end{tabular}%
}

\end{table*}

\subsection{Main Results}
\label{sec:main-results}

We compare the accuracy and token efficiency of \name against all fixed single-modality, two-modality, and full-modality interaction baselines across eight benchmarks.

\paragraph{Advantage over Fixed Modality Interactions.}
As shown in Table~\ref{tab:main-results}, \name achieves the highest accuracy on all eight datasets with competitive or even lower token costs in most cases. This gain is twofold. 
First, single-modality baselines suffer from under-fusion, and the most accurate fixed multimodal combination improves accuracy over the strongest single-modality baseline ($\mathrm{X}$-only) by up to 6.4\%. Second, \name outperforms the most accurate fixed combination on all benchmarks, improving accuracy by up to 4.2\%. At the other extreme, the full combination $\mathrm{T+G+X}$ suffers from over-fusion, with higher token costs but lower accuracy than \name across all eight benchmarks. Specifically, \name reduces token cost by 14.1\%-54.3\% alongside accuracy improvements. 

Critically, we find that \emph{no single fixed-modality combination is universally dominant}. The combination $\mathrm{T+X}$ achieves the best fixed-combination accuracy on GQA (77.6\%), VQA-v2 (84.0\%), DCASE (50.9\%), and IntentQA (71.2\%), while $\mathrm{G+X}$ leads on VCR (81.1\%), and $\mathrm{T+G+X}$ leads on OKVQA (93.5\%), Clotho (45.1\%), and NextQA (77.2\%). \name outperforms each of these dataset-specific best combinations, demonstrating its ability to adapt the modality combination to the evidence requirements of each instance rather than relying on a fixed choice. These gains follow from the accuracy-first objective of \name, which prioritizes answer accuracy and then minimizes token cost among the combinations expected to answer correctly. For example, on OKVQA, the most accurate fixed combination is $\mathrm{T+G+X}$ with 93.5\% accuracy and 9,773 tokens. \name reaches a higher accuracy of 93.8\% while using only 6,775 tokens (30.7\% fewer). These results remain consistent across three inference runs and five independent router retraining runs with different random seeds. 

\paragraph{Cross-Task Analysis.}
While under-fusion and over-fusion occur consistently across all three task families, the token cost of the task-dependent configuration $\mathrm{X}$ and the relative contribution of each modality differ by task. The token cost of $\mathrm{X}$ varies substantially, from comparable to $\mathrm{T}$ in VQA (e.g., 6,455 for $\mathrm{X}$ vs. 6,615 for $\mathrm{T}$ on GQA) and Audio-QA (e.g., 4,268 for $\mathrm{X}$ vs. 3,692 for $\mathrm{T}$ on Clotho), to over 100K tokens per instance in Video-QA (e.g., 109,732 on IntentQA). This cost structure motivates how \name selects the modality combination. In VQA and Audio-QA, where $\mathrm{X}$ provides decisive evidence at reasonable token cost, \name frequently includes $\mathrm{X}$. In Video-QA, $\mathrm{X}$ remains valuable for accuracy (e.g., 68.4\% for $\mathrm{X}$-only vs. 64.0\% for $\mathrm{T}$-only on IntentQA), but its token cost is much higher. Therefore, \name includes $\mathrm{X}$ only when temporal evidence is decisive to achieve the higher accuracy while reducing token cost by over 50\% compared to $\mathrm{T+G+X}$. Additionally, the contribution of $\mathrm{G}$ is also task-dependent: while $\mathrm{G}$-only is the weakest baseline in VQA, it performs more competitively in Audio-QA. These results highlight that none of the modalities is universally effective or redundant across tasks, and thus our instance-level adaptive routing is necessary.

\subsection{FaithScore Analysis}
\label{sec:faithscore}

\begin{table}[t]
\centering
\caption{FaithScore comparison across eight multimodal QA benchmarks.
Higher FaithScore indicates better-grounded answers with less hallucination.}
\label{tab:faithscore-underfusion}
\vspace{0.5em}
\begin{tabular}{lcccccccc}
\toprule
\multicolumn{1}{c}{\multirow{2.5}{*}{\textbf{Dataset}}}
& \multicolumn{3}{c}{\textbf{One Modality}}
& \multicolumn{3}{c}{\textbf{Two Modalities}}
& \multicolumn{1}{c}{\textbf{All}}
& \multicolumn{1}{c}{\textbf{Adaptive}} \\
\cmidrule(lr){2-4} \cmidrule(lr){5-7} \cmidrule(lr){8-8} \cmidrule(lr){9-9}
& $\mathrm{T}$ & $\mathrm{G}$ & $\mathrm{X}$ & $\mathrm{T+G}$ & $\mathrm{T+X}$ & $\mathrm{G+X}$ & $\mathrm{T+G+X}$ & \name \\
\midrule
GQA    & 0.61 & 0.50 & 0.64 & 0.64 & 0.71 & 0.66 & 0.69 & \textbf{0.72} \\
OKVQA  & 0.67 & 0.64 & 0.76 & 0.78 & 0.80 & 0.76 & 0.80 & \textbf{0.81} \\
VCR    & 0.63 & 0.66 & 0.66 & 0.68 & 0.68 & \textbf{0.75} & \textbf{0.75} & \textbf{0.75} \\
VQA-v2 & 0.71 & 0.63 & 0.74 & 0.77 & \textbf{0.78} & 0.74 & 0.77 & 0.77 \\ \midrule
Clotho    & 0.76 & 0.69 & 0.77 & 0.78 & 0.82 & \textbf{0.87} & 0.81 & 0.84 \\
DCASE & 0.67 & 0.64 & 0.67 & 0.68 & \textbf{0.70} & 0.68 & 0.69 & \textbf{0.70} \\ \midrule
IntentQA    & 0.82 & 0.81 & 0.83 & \textbf{0.84} & \textbf{0.84} & 0.83 & 0.82 & \textbf{0.84} \\
NextQA & 0.84 & 0.82 & 0.86 & 0.87 & 0.86 & \textbf{0.88} & \textbf{0.88} & \textbf{0.88} \\
\bottomrule
\end{tabular}
\end{table}

To evaluate the severity of under-fusion across modality configurations, we measure FaithScore, which quantifies how well each claim in a generated answer is grounded in the input evidence. The higher scores indicate fewer hallucinated claims~\cite{jing-etal-2024-faithscore}. We use FaithScore to detect under-fusion, since insufficient modality evidence forces agents to fill evidence gaps with ungrounded reasoning, which lowers the score. As shown in~Table~\ref{tab:faithscore-underfusion}, on GQA, the FaithScore of $\mathrm{G}$ and $\mathrm{T}$ is 0.50 and 0.61, respectively, far below the 0.72 of \name. This indicates that the agents lack grounding and hallucinate, which degrades accuracy. However, FaithScore penalizes any claim not directly grounded in the input, including correct inferences from commonsense or causal reasoning. Sufficient grounding is necessary to avoid harmful hallucination, but maximizing FaithScore beyond that level is undesirable, as it discourages productive inference~\cite{yang2026breaking}. This explains why \name achieves a FaithScore of 0.77 on VQA-v2, marginally below $\mathrm{T+X}$ (0.78), yet obtains much higher accuracy (87.7\% vs. 84.0\%). 

FaithScore distributions also vary across task families. VQA and Audio-QA vary widely across combinations (e.g., 0.50-0.72 on GQA and 0.69-0.87 on Clotho). In other words, on these tasks, which modality combination the agents use largely decides how well their answers are grounded in the evidence. Video-QA shows uniformly high scores (e.g., 0.82-0.88 on NextQA), since video combines visual, temporal, and audio information, leaving less room for hallucination. Overall, \name achieves the highest or tied-highest FaithScore on six of the eight datasets, consistently avoiding under-fusion. On the other two datasets, \name's FaithScore is marginally below the per-dataset best but yields higher accuracy, reflecting productive inference rather than evidence insufficiency.
We provide case studies where FaithScore penalizes correct inferences of \name (see Appendix~\ref{sec:faithscore-cases}).

\subsection{Comparison with Single-group Adaptive Routers}
\label{sec:routing-analysis}

\begin{table}[t]
\centering
\caption{Accuracy (Acc) and average token cost (Tkn) of \name and four single-group adaptive routers, each of which uses only one type of routing feature. Accuracy is reported as a percentage (\%), and Tkn denotes the average per-instance token cost, where K denotes thousands.}
\label{tab:routing-analysis}
\vspace{0.5em}
\setlength{\tabcolsep}{5pt}
\begin{tabular}{lcccccccccc}
\toprule
\multicolumn{1}{c}{\multirow{2.5}{*}{\textbf{Dataset}}}
& \multicolumn{2}{c}{\textbf{Question Type}}
& \multicolumn{2}{c}{\textbf{Confidence}}
& \multicolumn{2}{c}{\textbf{Missing Evidence}}
& \multicolumn{2}{c}{\textbf{Cost-Aware}}
& \multicolumn{2}{c}{\textbf{\name}} \\
\cmidrule(lr){2-3} \cmidrule(lr){4-5} \cmidrule(lr){6-7} \cmidrule(lr){8-9} \cmidrule(lr){10-11}
& \textbf{Acc} & \textbf{Tkn}
& \textbf{Acc} & \textbf{Tkn}
& \textbf{Acc} & \textbf{Tkn}
& \textbf{Acc} & \textbf{Tkn}
& \textbf{Acc} & \textbf{Tkn} \\
\midrule
GQA      & 78.9 & 6,508 & 78.3 & 6,634 & 78.7 & 6,673 & 78.1 & \textbf{6,102} & \textbf{80.1} & 6,334 \\
OKVQA    & 93.5 & 6,897 & 93.2 & 6,767 & 93.4 & 6,994 & 93.5 & \textbf{6,433} & \textbf{93.8} & 6,775 \\
VCR      & 83.7 & 8,417 & 82.1 & 9,332 & 82.4 & 8,937 & 81.6 & 7,943 & \textbf{84.2} & \textbf{7,788} \\
VQA-v2   & 86.4 & 7,570 & 85.2 & 8,434 & 86.1 & 8,892 & 85.6 & 7,789 & \textbf{87.7} & \textbf{7,006} \\
\midrule
Clotho   & 46.2 & 3,891 & 45.7 & \textbf{3,706} & 46.0 & 3,809 & 44.1 & 3,778 & \textbf{46.8} & 3,795 \\
DCASE    & 49.7 & 3,987 & 50.9 & 4,382 & 51.1 & 4,208 & 50.5 & 3,902 & \textbf{51.8} & \textbf{3,891} \\
\midrule
IntentQA & 72.2 & 55.9K & 71.0 & 45.8K & 71.6 & 53.7K & 70.2 & \textbf{44.9K} & \textbf{75.4} & 45.8K \\
NextQA   & 78.7 & 49.1K & 77.9 & 48.6K & \textbf{78.8} & 52.5K & 74.9 & \textbf{44.7K} & \textbf{78.8} & 48.6K \\
\bottomrule
\end{tabular}%
\end{table}

To evaluate whether all four groups of routing features are necessary for routing, we compare \name with four single-group adaptive routers, each of which selects the modality combination from only one type of routing feature.
Question-type routing decides from what the question asks, confidence-only routing from how confident the baseline agent is in the initial answer, missing-evidence-only routing from what evidence the baseline agent reports missing and whether the task-dependent input contains this evidence, and the cost-aware heuristic from how much new evidence each modality would add. Therefore, each single-group adaptive router uses only part of the routing features, whereas \name uses all four of them. We provide the exact routing features of each single-group adaptive router in Appendix~\ref{subsec:simple-details}. For a fair comparison, all routers share the same auxiliary training datasets, LLM backbone, candidate modality combinations, and MAD protocol. 

Comparing Table~\ref{tab:routing-analysis} with Table~\ref{tab:main-results}, we find that all four single-group adaptive routers achieve higher average accuracy over the eight benchmarks than the best fixed combination $\mathrm{T+X}$, by 1.0\% on average. Hence, adaptive modality selection is beneficial even with one type of routing feature. However, one type of routing feature is still insufficient, since \name achieves higher or equal accuracy than every single-group adaptive router on all eight benchmarks. Even compared with question-type routing, the most accurate single-group adaptive router, \name improves the average accuracy by 1.16\% and reduces the average token cost by 8.6\%. Moreover, the cost-aware heuristic uses fewer tokens than \name on five benchmarks but results in 2.51\% lower average accuracy, which indicates under-fusion. Therefore, the accuracy of \name comes from using all four groups of routing features rather than from any single group. 
We further evaluate the robustness of \name and the four single-group adaptive routers under three types of routing-feature perturbations: flipping, masking, and shifting (see Appendix~\ref{subsec:robust-simple}). Specifically, we compare the resulting accuracy degradation across the methods.

\section{Discussion}
\label{sec:discussion}

Although \name improves accuracy while remaining token-efficient across VQA, Audio-QA, and Video-QA, it requires a one-time offline training phase per task family. Specifically, training the modality router requires collecting outcomes from all seven modality combinations on auxiliary datasets, which runs the full MAD pipeline seven times per training instance. Despite this upfront cost, the resulting router is lightweight and generalizes to unseen evaluation datasets without additional data collection. This cost could be further reduced through online adaptation. 

\name can also be extended in two directions. First, adaptive modality interaction may benefit broader multi-agent paradigms beyond MAD, such as collaborative software development~\cite{hong2023metagpt,qian-etal-2024-chatdev} and multi-agent document understanding~\cite{han2025mdocagent}, where fixed interaction modalities may also cause under-fusion or over-fusion. 
Second, since current LLM APIs count tokens based on both video duration and frame resolution, jointly optimizing temporal sampling and spatial compression alongside modality selection could further reduce token cost without sacrificing accuracy.

\section{Conclusion}
\label{sec:conclusion}

We presented \name, an adaptive multimodal MAD framework that selects the best interaction modality combination per instance to improve reasoning accuracy while remaining token-efficient. By generating a structured baseline response, extracting interpretable routing features, and applying a lightweight router, \name avoids both the risks of under-fusion and over-fusion, and thus provides agents with decisive evidence without redundant modalities. Across eight benchmarks spanning three multimodal task families, \name achieves the highest accuracy. Combining modalities improves accuracy by up to 6.4\% over the best single-modality baseline, and adapting the combination per instance adds up to 4.2\% over the most accurate fixed combination of each benchmark with up to 53.4\% fewer tokens. Future work will extend adaptive modality interaction to broader multi-agent paradigms and jointly optimize temporal and spatial compression for video.

\section*{Acknowledgments}
We sincerely thank the anonymous reviewers of NeurIPS 2026 for their constructive comments and suggestions. This research was supported in part by the Commonwealth Cyber Initiative (CCI) and the National Research Foundation of Korea (NRF) grant funded by the Korea government (MSIT) (No. RS-2026-25602748).

\bibliographystyle{unsrtnat}
\bibliography{0main}
\clearpage
\appendix
\section*{Appendix}

In the appendix, we include implementation details, additional analyses, and additional evaluation results. First, we provide the implementation details in Appendix~\ref{sec:impl-details} and describe the routing features with the SHAP-based feature importance analysis in Appendix~\ref{sec:shap-analysis}. Then, we extend the best combination analysis to Audio-QA and Video-QA in Appendix~\ref{sec:oracle-distribution-appendix} and analyze the latency and cost of \name in Appendix~\ref{sec:cost-analyses}. Next, we report the ablation studies in Appendix~\ref{sec:ablation-studies}, the cross-backbone results using GPT-5.4-nano in Appendix~\ref{sec:gpt-backbone}, the robustness analysis in Appendix~\ref{sec:robustness}, and the comparison with single-group adaptive routers in Appendix~\ref{sec:simple-routing}. Lastly, we provide FaithScore case studies in Appendix~\ref{sec:faithscore-cases}.

\section{Implementation Details}
\label{sec:impl-details}

In this section, we provide the implementation details for our experiments, including LLM agent configuration, the MAD protocol, graph interaction details, dataset descriptions, router configurations, feature extraction settings, temporal compression procedures, and result stability analysis. 

\paragraph{LLM Agent Configuration.}
For LLM agents, we use a temperature of 0.0 to reduce randomness in the outputs and a maximum of 512 tokens. The structured baseline agent proposes up to $K{=}3$ task-dependent modality summaries per instance. All compared methods share the same backbone, graph pipeline, crop tool, and debate protocol, varying only the interaction modality combination.

\paragraph{MAD Protocol.}
All methods use the same MAD protocol with three agents in affirmative, negative, and judge roles for at most five rounds. The affirmative and negative agents present and critique candidate answers. The debate stops when the judge reaches a conclusion. Otherwise, the judge must produce a final answer at the fifth round. The judge's answer is used for evaluation. This protocol is shared by \name and all fixed-modality baselines.

\paragraph{Graph Interaction.}
Graph interaction follows the blueprint graph design~\cite{zheng2024bdog}. \name and all fixed baselines use the same graph construction, pruning settings, and graph representation. The graph provides structured evidence. The original blueprint graph design supports only VQA, where the graph is built from entities and relations in the image. We extend the graph modality to Audio-QA and Video-QA by building the graph from sound events in the audio and from visible events or actions in the video, respectively.
For Audio-QA and Video-QA, the prompts serialize each graph as a JSON object with an \texttt{events} list and a \texttt{relationships} list. Each event has an identifier, a description, and attributes, where an event describes a sound event for Audio-QA and a visible event or action for Video-QA. Each relationship links two event identifiers and specifies the relation between the events.

\paragraph{Dataset Details.}
We evaluate on eight benchmarks across three task families. For VQA, GQA~\cite{gqa} evaluates compositional visual reasoning over real-world images using scene-graph annotations, OKVQA~\cite{marino2019okvqa} requires external knowledge, VCR~\cite{zellers2019vcr} evaluates visual commonsense reasoning over movie scenes, and VQA-v2~\cite{balanced_vqa_v2} contains natural image questions with balanced answer distributions. For Audio-QA, Clotho~\cite{lipping2022clotho} contains open-ended questions over environmental sound recordings, and DCASE~\cite{heaudio} focuses on acoustic event detection recordings. For Video-QA, IntentQA~\cite{li2023intentqa} evaluates intent and causal reasoning over daily-life videos, and NextQA~\cite{xiao2021next} evaluates temporal and causal dynamics in short video clips. Each router is trained on auxiliary instances disjoint from these benchmarks, with 500 instances from Visual7W~\cite{visual7W} and CLEVR~\cite{CLEVR} for VQA, 250 from AudioMCQ~\cite{he2025measuring} for Audio-QA, and 250 from Video-MME~\cite{fu2025video} for Video-QA.

\paragraph{Router Configurations.}
The adaptive modality router comprises a correctness classifier and a token-cost regressor, with task-specific configurations for each task family. For VQA, the correctness classifier is a CatBoost~\cite{prokhorenkova2018catboost} model trained with binary cross-entropy~\cite{BCE} loss, using 380 estimators, a maximum tree depth of 4, and a learning rate of 0.03. The token-cost regressor is a LightGBM~\cite{lightgbm} model trained with root mean squared error~\cite{RMSE} loss, using 330 estimators, a maximum tree depth of 4, and a learning rate of 0.03. The acceptance threshold is $\tau = 0.52$, and the under-fusion pruning margin is $\Delta_{\mathrm{UF}} = 0.08$.
For Audio-QA and Video-QA, both the correctness classifier and token-cost regressor are XGBoost~\cite{chen2016xgboost} models. The classifier uses 300 estimators, a maximum tree depth of 6, and a learning rate of 0.05; the regressor uses 300 estimators, a maximum tree depth of 5, and a learning rate of 0.05. Meanwhile, the acceptance threshold is $\tau = 0.55$, and the under-fusion pruning margin is $\Delta_{\mathrm{UF}} = 0.10$. These hyperparameters are selected via grid search on the validation set of the training data and remain fixed across all evaluation benchmarks.

\paragraph{Feature Extraction Details.}
We extract 109 routing features for VQA, 98 for Audio-QA, and 100 for Video-QA. For feature extraction, we use all-MiniLM-L6-v2~\cite{reimers-2019-sentence-bert} for text embeddings and CLIP ViT-B/32~\cite{clip} for image embeddings, and each text or image representation is reduced to 32 dimensions by PCA~\cite{pca}. For VQA, we embed the full image and each of the up to three proposed regions. We then aggregate the region embeddings into two representations, one by taking the mean over the regions and one by taking the maximum over the regions in each dimension, and compare each representation with the full image embedding to obtain the image vs.\ crop-mean and image vs.\ crop-max similarity features. For Audio-QA, we use wav2vec2-base~\cite{baevski2020wav2vec} to encode the full recording and up to three proposed segments at 16 kHz, and a shared PCA projection reduces each representation to 16 dimensions. For Video-QA, we use X3D-M~\cite{feichtenhofer2020x3d} to encode the video and reduce each representation to 32 dimensions by PCA. We cap concatenated text at 600 words and use batch sizes of 64, 32, and 4 for text, image, and audio encoding, respectively. These embeddings are combined with the numerical and categorical routing features, and therefore the number of routing features differs from the input dimension of the correctness classifier and the token-cost regressor. The router models are trained on one NVIDIA RTX 4090 GPU.

\paragraph{Temporal Compression.}
For temporal compression, we apply the following procedure to compress routed evidence. For Audio-QA, \name compresses the selected audio segment through SciPy-based audio resampling~\cite{gommers2024scipy}. For Video-QA, \name compresses the visual stream through FFmpeg-based frame subsampling~\cite{newmarch2017ffmpeg}, compresses the accompanying audio track through SciPy-based audio resampling, and then recombines the two streams into a synchronized clip. The compression ratio $r$ is selected separately for each evaluation dataset: we use $r = 2.5$ for Clotho and IntentQA, and $r = 1$ (no compression) for DCASE and NextQA, reflecting that the latter two datasets contain content that is already sufficiently dense to leave no room for further compression.

\paragraph{Result Stability.}
To evaluate the stability of the results of \name, we consider two sources of variation on all eight benchmarks. First, we repeat the end-to-end evaluation three times with the same trained router and decoding settings, where each run includes the structured baseline LLM response, feature extraction, router selection, and final MAD execution. Since feature extraction and routing are deterministic given fixed LLM outputs, the variation across the three runs stems from minor nondeterminism in LLM API outputs even at temperature 0.0. The results of \name in Table~\ref{tab:main-results} are the mean over these three runs. Second, we retrain the router of each task family five times with different random seeds while keeping the auxiliary training data, routing features, hyperparameters, and LLM outputs fixed. From Table~\ref{tab:stability}, we find that the results of \name are stable under both sources of variation. Across the three inference runs, the standard deviation of accuracy is at most 0.16\% and the relative standard deviation of token cost is at most 3.78\%. Similarly, across the five retrainings, the standard deviation of accuracy is at most 0.35\% and the relative standard deviation of token cost is at most 4.74\%. Moreover, the mean accuracy over the five retrainings remains above the most accurate fixed combination on all eight benchmarks. Therefore, the conclusions in Section~\ref{sec:main-results} do not depend on one particular inference run or one particular random seed.
 
\begin{table}[t]
\centering
\caption{Stability of \name across three inference runs with the same trained router and across retraining the router with five different random seeds. Each cell reports the mean and one standard deviation.}
\label{tab:stability}
\vspace{0.5em}
\begin{tabular}{lcccc}
\toprule
\multicolumn{1}{c}{\multirow{2.5}{*}{\textbf{Dataset}}}
& \multicolumn{2}{c}{\textbf{Inference (3 runs)}}
& \multicolumn{2}{c}{\textbf{Retraining (5 seeds)}} \\
\cmidrule(lr){2-3} \cmidrule(lr){4-5}
& \textbf{Acc} & \textbf{Tkn} & \textbf{Acc} & \textbf{Tkn} \\
\midrule
GQA      & 80.1 $\pm$ 0.16 & 6,334 $\pm$ 96     & 80.3 $\pm$ 0.35 & 6,291 $\pm$ 183 \\
OKVQA    & 93.8 $\pm$ 0.03 & 6,775 $\pm$ 87     & 93.6 $\pm$ 0.21 & 6,832 $\pm$ 149 \\
VCR      & 84.2 $\pm$ 0.08 & 7,788 $\pm$ 161    & 83.9 $\pm$ 0.33 & 7,843 $\pm$ 281 \\
VQA-v2   & 87.7 $\pm$ 0.07 & 7,006 $\pm$ 122    & 87.7 $\pm$ 0.15 & 7,112 $\pm$ 282 \\
\midrule
Clotho   & 46.8 $\pm$ 0.08 & 3,795 $\pm$ 45     & 46.9 $\pm$ 0.24 & 3,701 $\pm$ 91 \\
DCASE    & 51.8 $\pm$ 0.09 & 3,891 $\pm$ 36     & 51.5 $\pm$ 0.13 & 3,848 $\pm$ 53 \\
\midrule
IntentQA & 75.4 $\pm$ 0.06 & 45,752 $\pm$ 1,010 & 75.2 $\pm$ 0.09 & 46,053 $\pm$ 1,880 \\
NextQA   & 78.8 $\pm$ 0.05 & 48,636 $\pm$ 1,840 & 79.0 $\pm$ 0.11 & 47,658 $\pm$ 2,260 \\
\bottomrule
\end{tabular}
\end{table}

\section{Feature Details and SHAP-based Feature Importance Analysis}
\label{sec:shap-analysis}

This section provides a comprehensive inventory of the routing features used by the modality router across all three task families and validates their contributions through SHapley Additive exPlanations (SHAP)~\cite{SHAP} analysis. The VQA router uses 109 features, the Audio-QA router uses 98 features, and the Video-QA router uses 100 features, reflecting differences in how each task-dependent modality $\mathrm{X}$ is represented in the feature space. Although the four-group organization is shared across all three task families, the third group adapts to the task-dependent modality $\mathrm{X}$ (image regions for VQA, audio segments for Audio-QA, spatio-temporal clips for Video-QA), and the fourth group adjusts the modality-specific gain and similarity features accordingly. We first detail how the features are organized into four groups aligned with the under-fusion and over-fusion routing risks introduced in Section~\ref{sec:method}. We then present per-feature SHAP importance results for both the correctness classifier and the token-cost regressor, confirming that all four feature groups contribute meaningfully and that no individual feature can be removed without losing useful routing signal.

\subsection{Router Feature Design}
\label{subsec:feature-design}

\paragraph{Feature Derivation.}
The routing features are derived from the question and the fields of the structured baseline output in two steps, which are shared across the three task families and require no additional LLM query and no evaluation labels. First, each feature group is derived from specific fields. The evidence requirement cues are derived from the question and the missing-evidence statement, and the uncertainty cues are derived from the initial answer, the confidence score, and the self-critique. Similarly, the evidence availability cues are derived from the candidate regions, which are image RoIs for VQA, audio segments for Audio-QA, and video clips for Video-QA, and the additional modality gain cues are derived from the overlap between each candidate modality and the structured baseline output. Second, each field is expanded by a small set of fixed operators. Numeric fields such as confidence and relevance scores are used directly, list fields such as target nouns are expanded into the list length and the leading elements, and lexical cues such as hedge terms and generic priors are matched into binary flags and counts. Moreover, each of the up to three candidate regions keeps separate features, and the relevance scores of the candidate regions are aggregated into the mean, standard deviation, minimum, maximum, the fraction of low-relevance regions, and a bucketed distribution. Finally, semantic overlap is measured by the cosine similarity between embeddings, and free-text fields are embedded and reduced by PCA. All derived features are kept without feature selection, since every feature contributes to at least one of the two router models in the SHAP analysis in Appendix~\ref{subsec:shap-results}.

\subsubsection{VQA Features (109 Features)}
\label{subsubsec:vqa-features}
 
Table~\ref{tab:feature_definitions} lists all 109 VQA features organized into four groups. The first three groups primarily detect \emph{under-fusion}, where the selected modality subset misses decisive evidence needed for a correct answer. The fourth group primarily reduces \emph{over-fusion}, where additional modalities contribute little useful evidence while unnecessarily increasing token cost.

\emph{Evidence requirement cues} (20 features) characterize the question's information needs by combining the question type and referenced entities with the baseline's missing-evidence statement. These features encode the extracted target nouns and their count, the question demand flags (e.g., whether answering requires spatial reasoning, OCR, counting, or interaction understanding), and a text statement of what visual evidence would confirm the hypothesis. They reveal the likely modality demand for each instance.

\emph{Uncertainty cues} (16 features) capture how well the baseline answer is supported. The core signals are the confidence score and the router's calibrated version of it, along with a calibration flag that labels each instance as overconfident, underconfident, or well-calibrated. Diagnostic signals detect hedge terms (e.g., \emph{maybe}, \emph{possibly}), generic priors (e.g., \emph{usually}, \emph{typically}), and whether the reasoning refers instance-specific visual evidence rather than relying on general knowledge. A self-critique field provides a counter-argument against the initial hypothesis. High uncertainty or reliance on generic priors can indicate hallucination risk, signaling that additional modalities may be needed.

\emph{Visual evidence availability cues} (52 features) estimate whether the missing evidence is likely present in the image. For each of up to three proposed image RoIs (crops), the features include the normalized bounding box coordinates, a confidence score that the crop contains decisive evidence, caption relevance to the question, and whether the caption mentions the target nouns. Aggregated statistics summarize the crop-level signals: mean, standard deviation, minimum, and maximum caption relevance; the fraction of low-relevance crops; the variance of crop confidence; and a relevance distribution. High-relevance crops that contain the key entities suggest that image-based reasoning can recover missing evidence, whereas low-relevance crops suggest that invoking the image modality is unlikely to help.

\emph{Additional modality gain cues} (21 features) estimate how much useful new evidence each candidate modality would contribute beyond the baseline. For the image-based modality, these include the estimated marginal gain in target-coverage, the overlap of text-image evidence, and the risk scores for irrelevant entities and distracting context. For the graph-based modality, the features capture the coverage of the target-node, the confidence of the subgraph, the quality of the relationship match and the risk scores for spurious relations, clutter and attribute loss. Seven cosine similarity features measure the semantic overlap between the question and various text artifacts (initial answer, self-critique, evidence statement, crop captions, and all artifacts combined) as well as between the full image and the crop embeddings. Low overlap between the question and existing artifacts suggests that adding a modality may introduce new information, while high overlap suggests less benefit from adding more modalities.

Within the 109 features, 71 are numeric values fed directly to the router models. The remaining 38 are free-text fields, captions, reasoning chains, and nested structured records that are PCA-embedded into fixed-dimensional vectors before routing. The numeric features are amenable to per-feature SHAP analysis; the PCA-embedded features contribute collectively through their principal components but cannot be attributed individually.

\subsubsection{Audio-QA Features (98 Features)}
\label{subsubsec:audio-features}

Table~\ref{tab:audio_feature_definitions} lists all 98 Audio-QA features. The following describes the differences between Audio-QA and VQA features.

\emph{Evidence requirement cues} (20 features) retain the same target-noun and demand-structure features as VQA but replace the six visual demand flags with audio demand flags. For example, whether the question requires understanding speech content, identifying specific sounds or instruments, temporal reasoning about the order of auditory events, recognizing emotion or tone, identifying speakers, or counting auditory entities.

\emph{Uncertainty cues} (16 features) are identical to VQA. Phase~{\large \textcircled{\small 1}} always begins with a text-only response regardless of the task family, so the uncertainty signals are the same across all three settings.

\emph{Audio evidence availability cues} (43 features) estimate whether the missing evidence is likely present in the audio signal. Instead of image crops with spatial bounding boxes, the structured prompt proposes up to three audio clips, each defined by a start and end timestamp. For each clip, the features include the temporal boundaries, a confidence score, a short description of the clip content, description relevance to the question, and whether the description mentions the target nouns. Aggregated statistics (mean, standard deviation, minimum, and maximum description relevance; fraction of low-relevance clips; variance of clip confidence; relevance distribution) summarize the clip-level signals. The number of features is less than VQA since audio segments only need two-coordinate temporal boundaries instead of four-coordinate bounding boxes across three proposed regions.

\emph{Additional modality gain cues} (19 features) follow the same structure as VQA but replace image-specific gain and risk features with audio-specific ones: estimated marginal coverage gain from the audio modality, text-audio evidence overlap, and risk scores for irrelevant and distracting audio content. The graph features are identical to VQA. Five cosine similarity features measure semantic overlap between the question and the text artifacts (initial answer, self-critique, evidence statement, clip descriptions, and all artifacts combined).

\subsubsection{Video-QA Features (100 Features)}
\label{subsubsec:video-features}

Table~\ref{tab:video_feature_definitions} lists all 100 Video-QA features. The following describes the differences between Video-QA and VQA features.

\emph{Evidence requirement cues} (20 features) use the same target-noun and demand-structure features as VQA but with six video-specific demand flags, which indicate whether the question requires action recognition, temporal reasoning about the sequence of events, spatial reasoning about entity relationships within frames, appearance recognition, OCR, or counting entities across frames.

\emph{Uncertainty cues} (16 features) are identical to VQA.

\emph{Video evidence availability cues} (43 features) estimate whether the missing evidence is likely present in the video signal. The structured prompt proposes up to three video region clips, each defined by a spatial bounding box and start and end timestamps. For each clip, the features include the temporal boundaries, a confidence score, a short description, description relevance, target-noun containment, and matched-noun lists. 

\emph{Additional modality gain cues} (21 features) replace image-specific features used in VQA with video-specific ones: estimated marginal coverage gain from the video modality, text-video evidence overlap, and risk scores for irrelevant entities and distracting context in the proposed clips. The graph features are still identical across all three task families. Five cosine similarity features measure semantic overlap between the question and text artifacts. Video PCA embeddings are enabled in the Video-QA router using X3D-M~\cite{feichtenhofer2020x3d} with 32 principal components, so two additional similarity features are available: video vs clip-mean embedding and video vs clip-max embedding.

\subsection{SHAP-based Feature Importance Analysis}
\label{subsec:shap-results}
 
\begin{figure*}[p]
\centering
\includegraphics[width=0.7\linewidth]{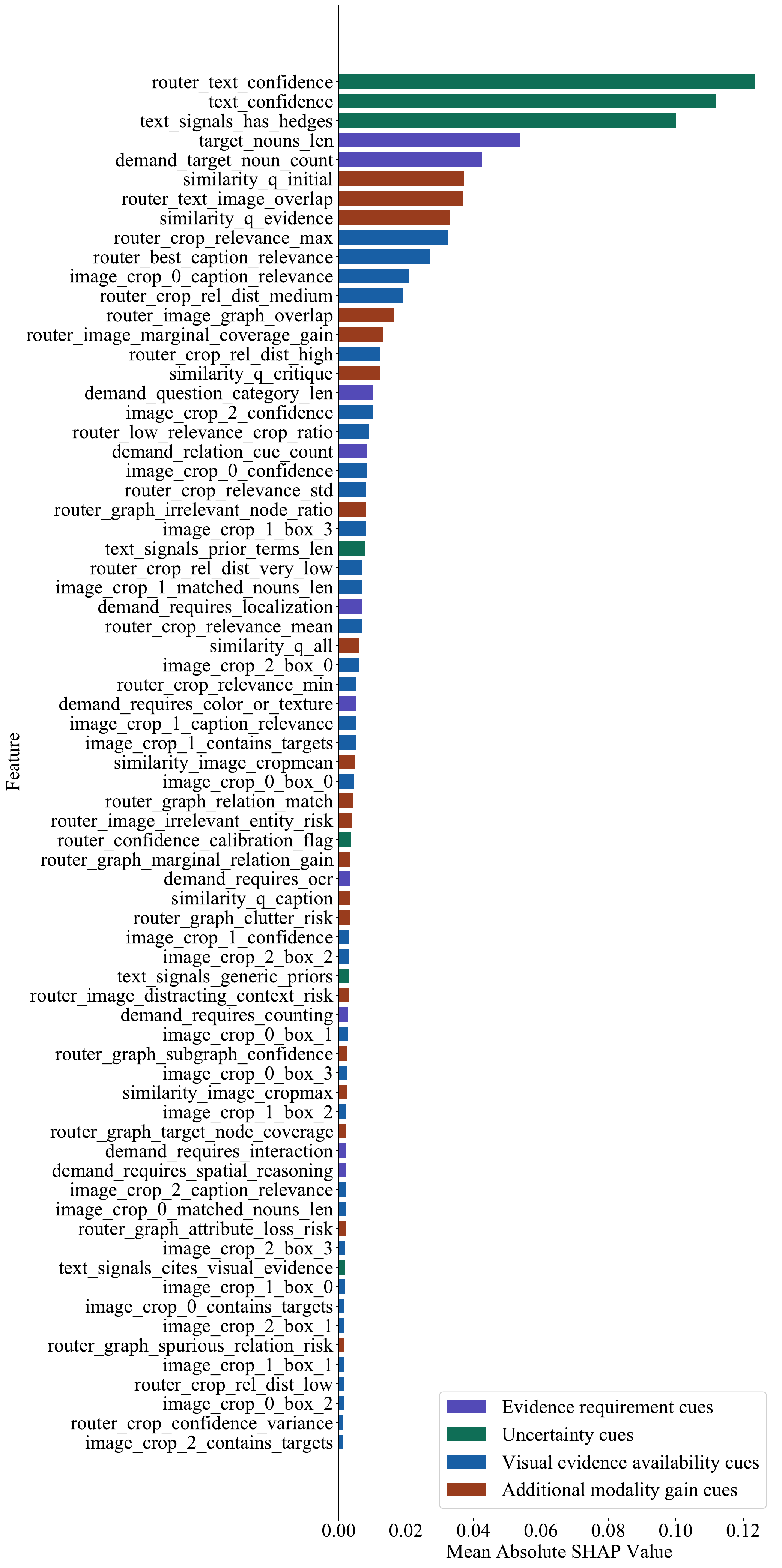}
\caption{Feature importance (mean absolute SHAP value) of 71 numeric VQA router features for the \textbf{correctness classifier} (CatBoost), which predicts whether the selected modality subset will yield a correct answer. The remaining 38 features (free-text fields, captions, reasoning chains, and nested structured records) are PCA-embedded before routing and excluded from per-feature SHAP analysis.}
\label{fig:shap-vqa-correctness}
\end{figure*}

\begin{figure*}[p]
\centering
\includegraphics[width=0.7\linewidth]{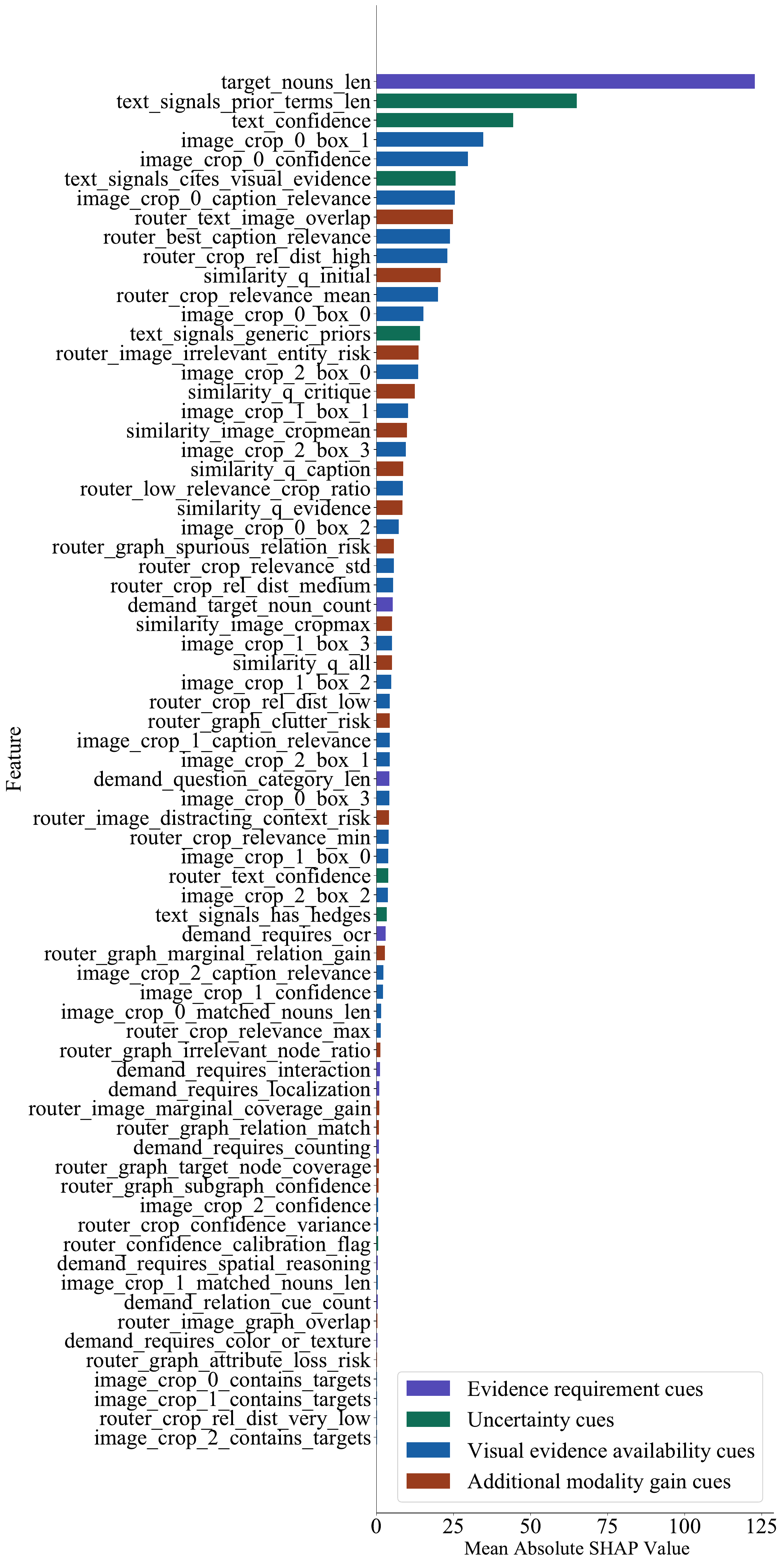}
\caption{Feature importance (mean absolute SHAP value) of 71 numeric VQA router features for the \textbf{token-cost regressor} (LightGBM).}
\label{fig:shap-vqa-cost}
\end{figure*}

\begin{figure*}[p]
\centering
\includegraphics[width=0.7\linewidth]{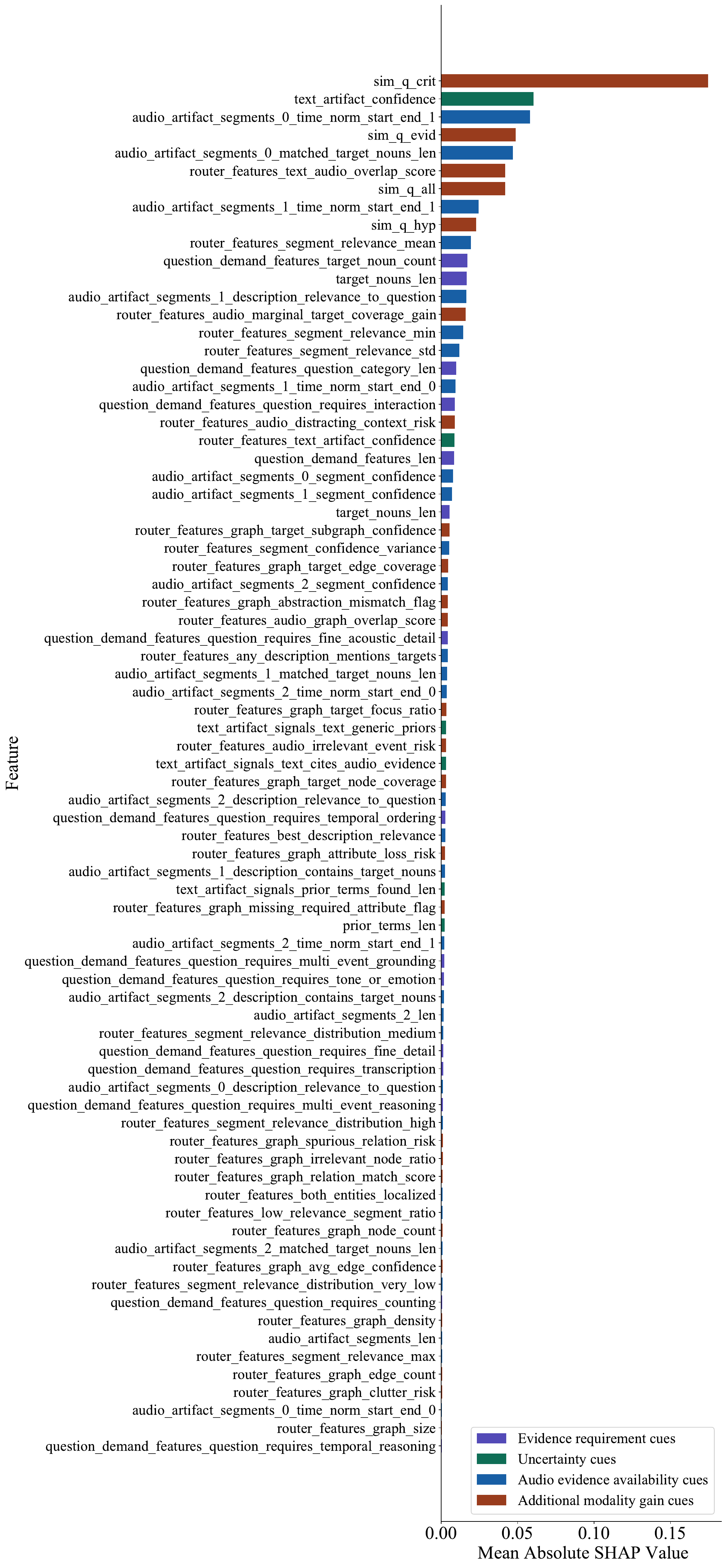}
\caption{Feature importance (mean absolute SHAP value) of 77 non-PCA Audio-QA router features for the \textbf{correctness classifier} (XGBoost).}
\label{fig:shap-audio-correctness}
\end{figure*}

\begin{figure*}[p]
\centering
\includegraphics[width=0.7\linewidth]{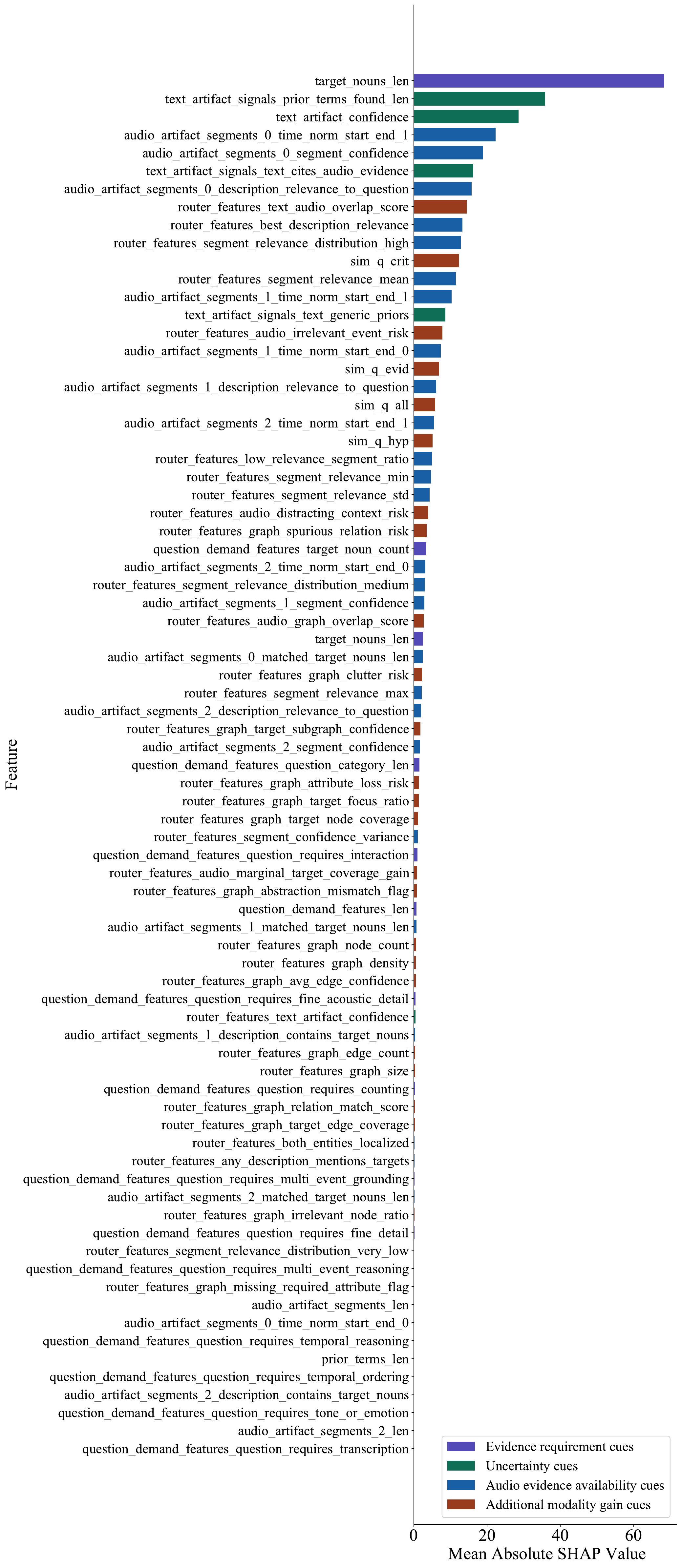}
\caption{Feature importance (mean absolute SHAP value) of 77 non-PCA Audio-QA router features for the \textbf{token-cost regressor} (XGBoost).}
\label{fig:shap-audio-cost}
\end{figure*}

\begin{figure*}[p]
\centering
\includegraphics[width=0.7\linewidth]{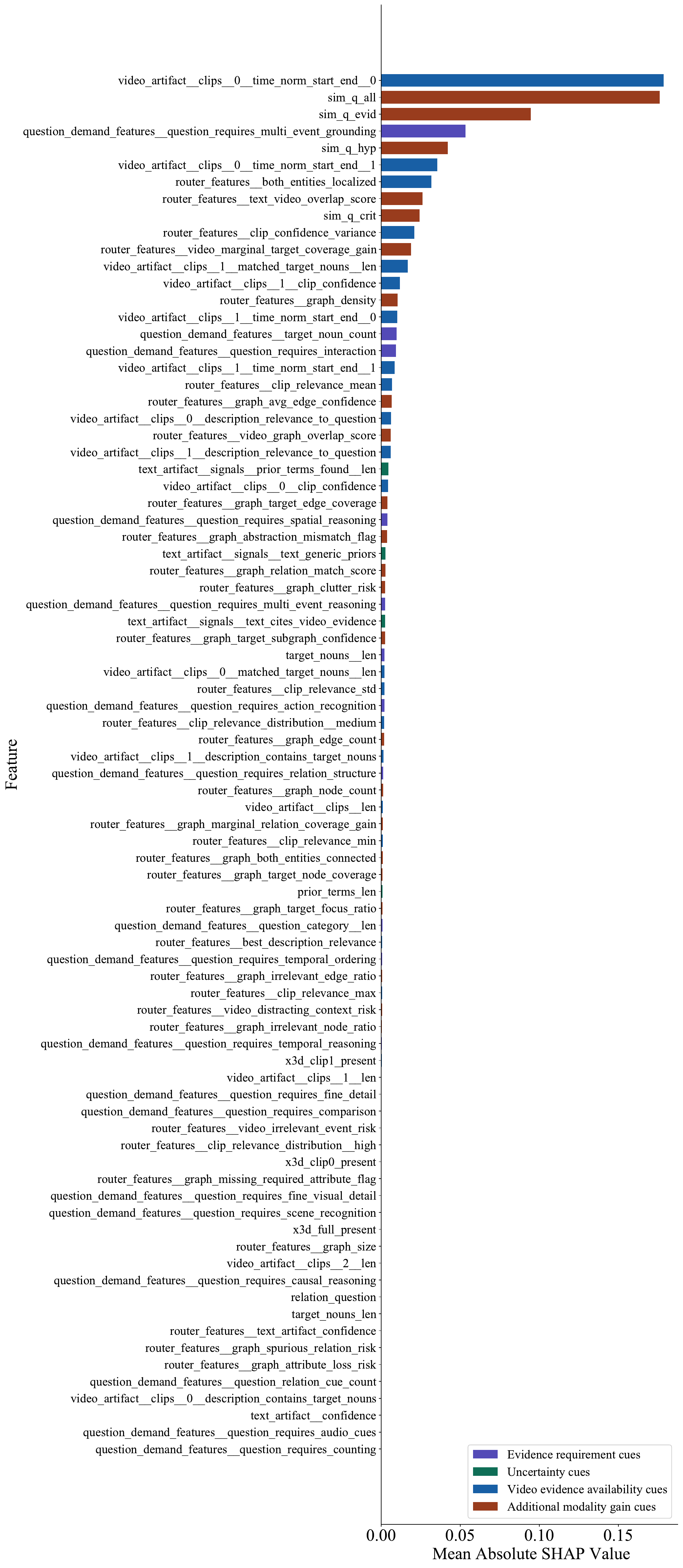}
\caption{Feature importance (mean absolute SHAP value) of 82 non-PCA Video-QA router features for the \textbf{correctness classifier} (XGBoost).}
\label{fig:shap-video-correctness}
\end{figure*}

\begin{figure*}[p]
\centering
\includegraphics[width=0.7\linewidth]{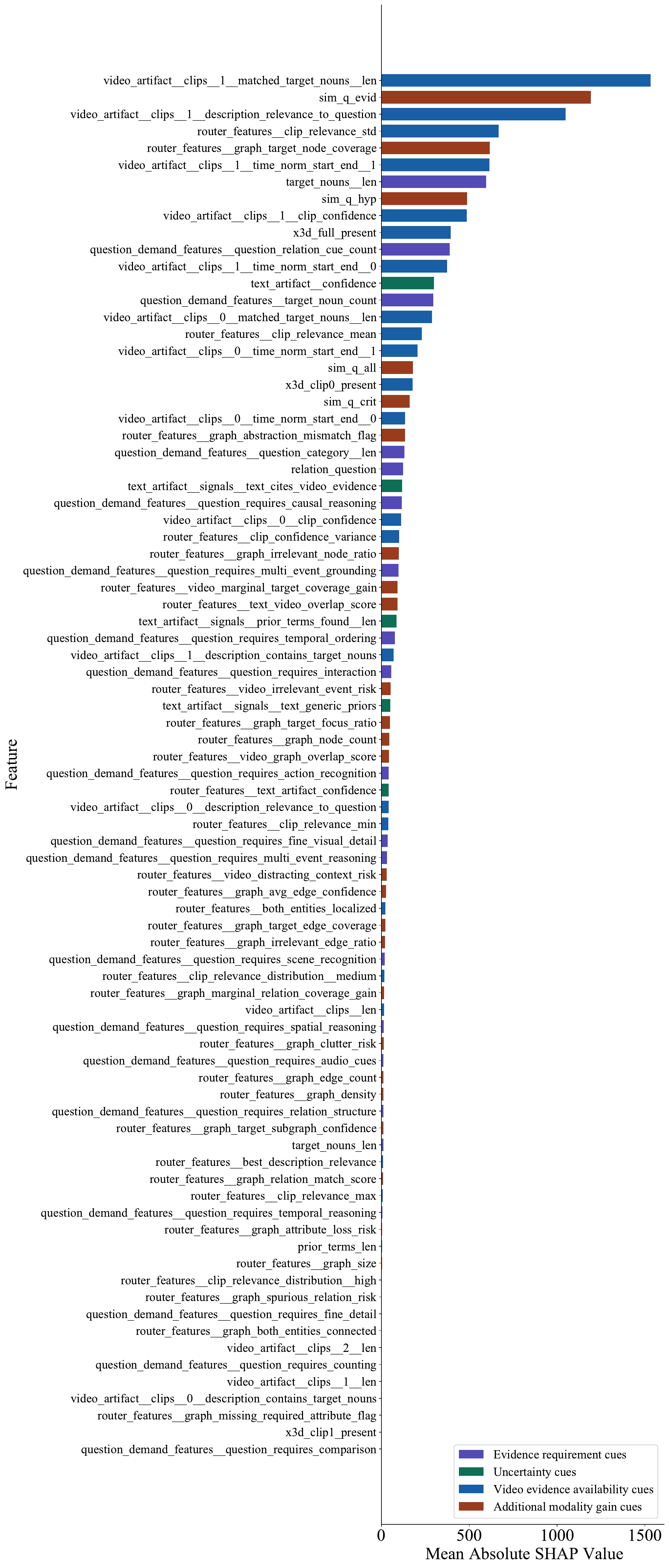}
\caption{Feature importance (mean absolute SHAP value) of 82 non-PCA Video-QA router features for the \textbf{token-cost regressor} (XGBoost).}
\label{fig:shap-video-cost}
\end{figure*}

To evaluate whether the routing features contribute meaningfully to the router's decisions across all three task families, we conduct a SHAP analysis on both the correctness classifier and the token-cost regressor for VQA, Audio-QA, and Video-QA. We compute SHAP values on the held-out test split using TreeExplainer~\cite{lundberg2020local}, which provides exact Shapley values for tree-based models, and report the mean absolute SHAP value for each numeric feature as the measure of global feature importance. The results are presented in Fig.~\ref{fig:shap-vqa-correctness} and~\ref{fig:shap-vqa-cost} for VQA (71 numeric features), Fig.~\ref{fig:shap-audio-correctness} and~\ref{fig:shap-audio-cost} for Audio-QA (77 numeric features), and Fig.~\ref{fig:shap-video-correctness} and~\ref{fig:shap-video-cost} for Video-QA (82 numeric features). Across all three task families, every numeric feature achieves a non-negligible mean absolute SHAP value in at least one of the two models, showing that each feature carries useful routing signal.

\paragraph{Correctness Classifier.}
The most important correctness signals differ across the three task families. 

For VQA, uncertainty cues are the most important group, contributing 38.1\% of the total SHAP importance. Text confidence (0.124) and hedge term indicator (0.100) showing that lexical markers of uncertainty (e.g., \emph{maybe}, \emph{possibly}) provide complementary evidence beyond the numeric confidence score. They help the router identify cases where the initial text answer may be unreliable. Visual evidence availability also contributes 25.3\%, showing that crop relevance and caption relevance help estimate whether image evidence can improve the answer. Additional modality gain cues contribute 21.7\%, while evidence requirement cues contribute 14.8\%, mainly through entity count features. 

In Audio-QA, additional modality gain cues take over at 48.3\%, driven by cosine similarity features that measure the semantic gap between the question and the baseline's reasoning. The question vs. self-critique similarity (0.175) is the single most important feature, with three more similarity features in the top seven. This shift aligns with the under-fusion tendency observed in Appendix~\ref{sec:ablation-cot}, since similarity features are the primary cues that distinguish audio instances requiring additional evidence from those already served well by text alone. Audio evidence availability cues contribute 31.8\% through segment temporal boundaries (0.058) and target noun matches (0.047), showing that the proposed audio segments carry discriminative routing signal. Uncertainty cues are less important than in VQA, contributing 9.8\%, while evidence requirement cues contribute 10.1\%.

In Video-QA, additional modality gain cues and video evidence availability contribute most to the correctness classifier, contributing 48.9\% and 39.8\%, respectively. The router relies heavily on whether the proposed clips are temporally relevant (0.179) and semantically aligned with the question (0.176), indicating that clip temporal position and semantic coverage are both critical for predicting correctness. Multi-event grounding (0.053) is also important, reflecting the need to reason over event sequences in video tasks. In contrast, uncertainty cues contribute only 1.3\%, showing that baseline text confidence is much less informative for Video-QA than for VQA.
 
Across the three settings, uncertainty cues decline sharply from 38.1\% in VQA to 9.8\% in Audio-QA and 1.3\% in Video-QA. As $\mathrm{X}$ moves from static images to temporal audio and video signals, the router shifts from relying on baseline confidence to relying on whether the proposed $\mathrm{X}$ regions contain decisive evidence.

\paragraph{Token Cost Regressor.}
For token cost prediction, modality-specific evidence availability is the strongest signal across all three task families. It contributes 39.6\% in VQA, 39.3\% in Audio-QA, and 52.0\% in Video-QA. This pattern shows that the cost model mainly depends on how much visual, audio, or video evidence needs to be processed. 
In VQA, visual crop features contribute most to token cost prediction, with the first crop's bounding box coordinate (34.6), crop confidence (29.7), and caption relevance (25.5) ranking among the top features. Evidence requirement cues are also important, with entity count (122.7) serving as the strongest individual cost predictor. Questions involving more target entities tend to require more crop descriptions and more relational reasoning. 
In Audio-QA, the cost pattern is similar, but the SHAP magnitudes are lower. Entity count (68.4) remains the strongest cost predictor, while audio segment boundaries and relevance scores provide additional cost signals. The lower SHAP scale matches the lower token cost range of Audio-QA. The group distribution is very similar to VQA, with uncertainty cues at 22.3\%, evidence requirement cues at 19.6\%, and gain cues at 18.8\%.
In Video-QA, video evidence availability has the largest impact on token cost. Features related to the second proposed clip are especially important, suggesting that deciding whether to use an additional clip is a major source of cost variation. Modality gain cues and evidence requirement cues also contribute strongly. The cost SHAP magnitudes are much larger than those in VQA and Audio-QA, consistent with the substantially higher token costs of Video-QA.

\paragraph{Cross-Task Comparison.} The SHAP results show that different task families require different routing signals. For correctness prediction, the main signal shifts from uncertainty in VQA to modality gain and task-dependent evidence availability in Audio-QA and Video-QA. This indicates that routing becomes more dependent on localized task-dependent evidence as the modality changes from images to audio and video. For token cost prediction, the three task families share a common pattern. Evidence availability and evidence requirement cues are consistently strong predictors. Entity count is especially useful across tasks, while modality-specific evidence features capture the cost of processing crops, segments, and clips. These results validate the four feature groups in \name: evidence requirements, baseline uncertainty, modality-specific evidence availability, and marginal modality gain. Each group provides complementary signals for accurate and token-efficient modality selection across tasks.

\clearpage
{\footnotesize
\setlength{\tabcolsep}{2.5pt}
\setlength{\LTpre}{0pt}
\setlength{\LTpost}{6pt}

\begin{xltabular}{\textwidth}{@{}p{1.8cm}lX@{}}

\caption{List of 109 router features organized by routing-risk group.
The first three groups primarily detect \emph{under-fusion} (missing decisive evidence); the fourth group primarily reduces \emph{over-fusion} (unnecessary token cost).}
\label{tab:feature_definitions} \\
\toprule
\textbf{Group} & \textbf{Feature Name} & \textbf{Description} \\
\midrule
\endfirsthead

\toprule
\textbf{Group} & \textbf{Feature Name} & \textbf{Description} \\
\midrule
\endhead

\midrule
\multicolumn{3}{r}{\itshape Continued on next page} \\
\endfoot
\bottomrule
\\[-6pt]

\endlastfoot

& \texttt{target\_nouns} & Key nouns/entities from question and options \\
& \texttt{target\_nouns\_len} & \# of extracted target nouns \\
& \texttt{target\_nouns\_\{0,1,2,3\}} & Target noun at index 0-3 \textbf{(4 features)} \\
& \texttt{demand} & Structured question demand summary \\
& \texttt{demand\_question\_category} & Assigned high-level question categories \\
& \texttt{demand\_question\_category\_len} & \# of assigned categories \\
& \texttt{demand\_question\_category\_\{0,1\}} & Category at index 0, 1 \textbf{(2 features)} \\
& \texttt{demand\_target\_noun\_count} & \# of target nouns referenced in question \\
\multirow{-9}{*}[-4em]{\shortstack[l]{\textbf{Evidence}\\\textbf{requirement}\\\textbf{cues}}}
  & \texttt{demand\_requires\_localization} & Requires locating entities in image \\
& \texttt{demand\_requires\_interaction} & Requires understanding an interaction \\
& \texttt{demand\_requires\_spatial\_reasoning} & Requires spatial reasoning \\
& \texttt{demand\_requires\_color\_or\_texture} & Requires color or texture cues \\
& \texttt{demand\_requires\_ocr} & Requires reading text (OCR) \\
& \texttt{demand\_requires\_counting} & Requires counting entities \\
& \texttt{demand\_relation\_cue\_count} & \# of relation/spatial cue terms in question \\
& \texttt{text\_evidence\_to\_confirm} & Visual evidence needed to confirm hypothesis \\

\midrule

& \texttt{text} & Structured text-only output \\
& \texttt{text\_initial\_answer} & short initial answer justification from text output \\
& \texttt{text\_confidence} & Text output confidence score [0-1] \\
& \texttt{text\_reasoning} & Short reasoning supporting the initial answer \\
& \texttt{text\_signals} & Diagnostic signals from text artifact \\
& \texttt{text\_signals\_has\_hedges} & Whether hedge terms (e.g., \emph{maybe}) appear \\
& \texttt{text\_signals\_hedge\_terms\_found} & List of detected hedge terms \\
& \texttt{text\_signals\_cites\_visual\_evidence} & Whether visual evidence is cited in reasoning \\
\multirow{-9}{*}[-4em]{\shortstack[l]{\textbf{Uncertainty}\\\textbf{cues}}}
  & \texttt{text\_signals\_generic\_priors} & Whether generic priors (e.g., \emph{usually}) are used \\
& \texttt{text\_signals\_prior\_terms\_found} & List of detected generic-prior terms \\
& \texttt{text\_signals\_prior\_terms\_len} & \# of generic-prior terms found \\
& \texttt{text\_signals\_prior\_terms\_0} & First generic-prior term found \\
& \texttt{text\_self\_critique} & Self-critique of the hypothesis answer \\
& \texttt{text\_self\_critique\_neg\_reasoning} & Counter-argument against the hypothesis \\
& \texttt{router\_text\_confidence} & Router's text-artifact confidence score \\
& \texttt{router\_confidence\_calibration\_flag} & Calibration: overconfident / underconfident / ok \\

\midrule

& \texttt{image} & Structured image-grounded output \\
& \texttt{image\_crops} & List of proposed image crops \\
& \texttt{image\_crop\_$k$} & Record for crop $k$, $k\!\in\!\{0,1,2\}$ \textbf{(3 features)} \\
& \texttt{image\_crop\_$k$\_box} & Normalized bounding box for crop $k$ \textbf{(3 features)} \\
& \texttt{image\_crop\_$k$\_box\_\{0..3\}} & Box coords x1,y1,x2,y2 for crop $k$ \textbf{(12 features)} \\
& \texttt{image\_crop\_$k$\_confidence} & Confidence crop $k$ has decisive evidence \textbf{(3 features)} \\
& \texttt{image\_crop\_$k$\_caption} & Short caption of crop $k$ \textbf{(3 features)} \\
& \texttt{image\_crop\_$k$\_caption\_relevance} & Caption relevance to question for crop $k$ \textbf{(3 features)} \\
& \texttt{image\_crop\_$k$\_contains\_targets} & Whether crop $k$ caption mentions targets \textbf{(3 features)} \\
& \texttt{image\_crop\_$k$\_matched\_nouns} & Target nouns matched in crop $k$ \textbf{(3 features)} \\
& \texttt{image\_crop\_\{0,1\}\_matched\_nouns\_len} & \# of matched target nouns in crop 0, 1 \textbf{(2 features)} \\
& \texttt{image\_crop\_\{0,1\}\_matched\_nouns\_0} & First matched noun for crop 0, 1 \textbf{(2 features)} \\
& \texttt{image\_crop\_0\_matched\_nouns\_1} & Second matched noun for crop 0 \\
\multirow{-14}{*}[-7em]{\shortstack[l]{\textbf{Visual}\\\textbf{evidence}\\\textbf{availability}\\\textbf{cues}}}
  & \texttt{router\_best\_caption\_relevance} & Max caption relevance across all crops \\
& \texttt{router\_crop\_relevance\_mean} & Mean caption relevance across crops \\
& \texttt{router\_crop\_relevance\_std} & Std.\ of caption relevance across crops \\
& \texttt{router\_crop\_relevance\_min} & Min caption relevance across crops \\
& \texttt{router\_crop\_relevance\_max} & Max caption relevance across crops \\
& \texttt{router\_low\_relevance\_crop\_ratio} & Fraction of crops with low relevance \\
& \texttt{router\_crop\_confidence\_variance} & Variance of crop confidence scores \\
& \texttt{router\_crop\_relevance\_distribution} & Bucketed crop relevance distribution \\
& \texttt{router\_crop\_rel\_dist\_very\_low} & Share of crops in very-low relevance bucket \\
& \texttt{router\_crop\_rel\_dist\_low} & Share of crops in low relevance bucket \\
& \texttt{router\_crop\_rel\_dist\_medium} & Share of crops in medium relevance bucket \\
& \texttt{router\_crop\_rel\_dist\_high} & Share of crops in high relevance bucket \\

\midrule

& \texttt{router} & Structured router feature bundle \\
& \texttt{router\_image\_marginal\_coverage\_gain} & Estimated target-coverage gain from image \\
& \texttt{router\_text\_image\_overlap} & Text-image evidence overlap \\
& \texttt{router\_image\_irrelevant\_entity\_risk} & Risk that crops emphasize irrelevant entities \\
\multirow{-11}{*}[-4em]{\shortstack[l]{\textbf{Additional}\\\textbf{modality}\\\textbf{gain cues}}}
  
& \texttt{router\_image\_distracting\_context\_risk} & Risk that crops include distracting context \\
& \texttt{router\_graph\_target\_node\_coverage} & Target-entity coverage among graph nodes \\
& \texttt{router\_graph\_subgraph\_confidence} & Confidence of target-focused subgraph \\
& \texttt{router\_graph\_relation\_match} & Graph-question relation match \\
& \texttt{router\_graph\_irrelevant\_node\_ratio} & Fraction of irrelevant graph nodes \\
& \texttt{router\_graph\_spurious\_relation\_risk} & Risk of spurious graph relations \\
\multirow{-11}{*}[-9em]{\shortstack[l]{\textbf{Additional}\\\textbf{modality}\\\textbf{gain cues}}}
  & \texttt{router\_graph\_clutter\_risk} & Risk of graph clutter or noise \\
& \texttt{router\_graph\_attribute\_loss\_risk} & Risk of attribute loss in graph abstraction \\
& \texttt{router\_graph\_marginal\_relation\_gain} & Relation-coverage gain from graph \\
& \texttt{router\_image\_graph\_overlap} & Image-graph evidence overlap \\
& \texttt{similarity\_q\_initial} & Cosine similarity: question vs.\ initial answer \\
& \texttt{similarity\_q\_critique} & Cosine similarity: question vs.\ self-critique \\
& \texttt{similarity\_q\_evidence} & Cosine similarity: question vs.\ evidence \\
& \texttt{similarity\_q\_caption} & Cosine similarity: question vs.\ crop captions \\
& \texttt{similarity\_q\_all} & Cosine similarity: question vs.\ all text artifacts \\
& \texttt{similarity\_image\_cropmean} & Cosine similarity: image vs.\ mean crop \\
& \texttt{similarity\_image\_cropmax} & Cosine similarity: image vs.\ max crop \\

\end{xltabular}
}

 
\clearpage
{\footnotesize
\setlength{\tabcolsep}{2.5pt}
\setlength{\LTpre}{0pt}
\setlength{\LTpost}{6pt}
 
\begin{xltabular}{\textwidth}{@{}p{1.8cm}lX@{}}

\caption{List of 98 Audio-QA router features organized by routing-risk group.
The first three groups primarily detect \emph{under-fusion} (missing decisive evidence); the fourth group primarily reduces \emph{over-fusion} (unnecessary token cost).}
\label{tab:audio_feature_definitions} \\

\toprule
\textbf{Group} & \textbf{Feature Name} & \textbf{Description} \\
\midrule
\endfirsthead
 
\toprule
\textbf{Group} & \textbf{Feature Name} & \textbf{Description} \\
\midrule
\endhead
 
\midrule
\multicolumn{3}{r}{\itshape Continued on next page} \\
\endfoot
\bottomrule
\\[-6pt]

\endlastfoot
 
& \texttt{target\_nouns} & Key nouns/entities from question and options \\
& \texttt{target\_nouns\_len} & \# of extracted target nouns \\
& \texttt{target\_nouns\_\{0,1,2,3\}} & Target noun at index 0--3 \textbf{(4 features)} \\
& \texttt{demand} & Structured question demand summary \\
& \texttt{demand\_question\_category} & Assigned high-level question categories \\
& \texttt{demand\_question\_category\_len} & \# of assigned categories \\
& \texttt{demand\_question\_category\_\{0,1\}} & Category at index 0, 1 \textbf{(2 features)} \\
& \texttt{demand\_target\_noun\_count} & \# of target nouns referenced in question \\
\multirow{-9}{*}[-4em]{\shortstack[l]{\textbf{Evidence}\\\textbf{requirement}\\\textbf{cues}}}
  & \texttt{demand\_requires\_speech\_content} & Requires understanding speech or dialogue \\
& \texttt{demand\_requires\_sound\_identification} & Requires identifying specific sounds or instruments \\
& \texttt{demand\_requires\_temporal\_reasoning} & Requires reasoning about temporal order of sounds \\
& \texttt{demand\_requires\_emotion\_or\_tone} & Requires recognizing emotion, mood, or tone \\
& \texttt{demand\_requires\_speaker\_identification} & Requires identifying who is speaking \\
& \texttt{demand\_requires\_counting} & Requires counting auditory entities or events \\
& \texttt{demand\_relation\_cue\_count} & \# of relation/temporal cue terms in question \\
& \texttt{text\_evidence\_to\_confirm} & Audio evidence needed to confirm hypothesis \\
 
\midrule
 
& \texttt{text} & Structured text-only output \\
& \texttt{text\_initial\_answer} & Short initial answer justification from text output \\
& \texttt{text\_confidence} & Text output confidence score [0--1] \\
& \texttt{text\_reasoning} & Short reasoning supporting the initial answer \\
& \texttt{text\_signals} & Diagnostic signals from text artifact \\
& \texttt{text\_signals\_has\_hedges} & Whether hedge terms (e.g., \emph{maybe}) appear \\
& \texttt{text\_signals\_hedge\_terms\_found} & List of detected hedge terms \\
& \texttt{text\_signals\_cites\_visual\_evidence} & Whether modality-specific evidence is cited \\
\multirow{-9}{*}[-4em]{\shortstack[l]{\textbf{Uncertainty}\\\textbf{cues}}}
  & \texttt{text\_signals\_generic\_priors} & Whether generic priors (e.g., \emph{usually}) are used \\
& \texttt{text\_signals\_prior\_terms\_found} & List of detected generic-prior terms \\
& \texttt{text\_signals\_prior\_terms\_len} & \# of generic-prior terms found \\
& \texttt{text\_signals\_prior\_terms\_0} & First generic-prior term found \\
& \texttt{text\_self\_critique} & Self-critique of the hypothesis answer \\
& \texttt{text\_self\_critique\_neg\_reasoning} & Counter-argument against the hypothesis \\
& \texttt{router\_text\_confidence} & Router's text-artifact confidence score \\
& \texttt{router\_confidence\_calibration\_flag} & Calibration: overconfident / underconfident / ok \\
 
\midrule
 
& \texttt{audio} & Structured audio-grounded output \\
& \texttt{audio\_clips} & List of proposed audio clips \\
& \texttt{audio\_clip\_$k$} & Record for clip $k$, $k\!\in\!\{0,1,2\}$ \textbf{(3 features)} \\
& \texttt{audio\_clip\_$k$\_start} & Start timestamp (seconds) for clip $k$ \textbf{(3 features)} \\
& \texttt{audio\_clip\_$k$\_end} & End timestamp (seconds) for clip $k$ \textbf{(3 features)} \\
& \texttt{audio\_clip\_$k$\_confidence} & Confidence clip $k$ has decisive evidence \textbf{(3 features)} \\
& \texttt{audio\_clip\_$k$\_description} & Short description of clip $k$ content \textbf{(3 features)} \\
& \texttt{audio\_clip\_$k$\_description\_relevance} & Description relevance to question for clip $k$ \textbf{(3 features)} \\
& \texttt{audio\_clip\_$k$\_contains\_targets} & Whether clip $k$ description mentions targets \textbf{(3 features)} \\
& \texttt{audio\_clip\_$k$\_matched\_nouns} & Target nouns matched in clip $k$ \textbf{(3 features)} \\
& \texttt{audio\_clip\_\{0,1\}\_matched\_nouns\_len} & \# of matched target nouns in clip 0, 1 \textbf{(2 features)} \\
& \texttt{audio\_clip\_\{0,1\}\_matched\_nouns\_0} & First matched noun for clip 0, 1 \textbf{(2 features)} \\
\multirow{-13}{*}[-7em]{\shortstack[l]{\textbf{Audio}\\\textbf{evidence}\\\textbf{availability}\\\textbf{cues}}}
  & \texttt{audio\_clip\_0\_matched\_nouns\_1} & Second matched noun for clip 0 \\
& \texttt{router\_best\_description\_relevance} & Max description relevance across all clips \\
& \texttt{router\_clip\_relevance\_mean} & Mean description relevance across clips \\
& \texttt{router\_clip\_relevance\_std} & Std.\ of description relevance across clips \\
& \texttt{router\_clip\_relevance\_min} & Min description relevance across clips \\
& \texttt{router\_clip\_relevance\_max} & Max description relevance across clips \\
& \texttt{router\_low\_relevance\_clip\_ratio} & Fraction of clips with low relevance \\
& \texttt{router\_clip\_confidence\_variance} & Variance of clip confidence scores \\
& \texttt{router\_clip\_relevance\_distribution} & Bucketed clip relevance distribution \\
& \texttt{router\_clip\_rel\_dist\_very\_low} & Share of clips in very-low relevance bucket \\
& \texttt{router\_clip\_rel\_dist\_low} & Share of clips in low relevance bucket \\
& \texttt{router\_clip\_rel\_dist\_medium} & Share of clips in medium relevance bucket \\
& \texttt{router\_clip\_rel\_dist\_high} & Share of clips in high relevance bucket \\
 
\midrule
 
& \texttt{router} & Structured router feature bundle \\
& \texttt{router\_audio\_marginal\_coverage\_gain} & Estimated target-coverage gain from audio \\
& \texttt{router\_text\_audio\_overlap} & Text--audio evidence overlap \\
& \texttt{router\_audio\_irrelevant\_content\_risk} & Risk that clips emphasize irrelevant content \\
\multirow{-5}{*}[-2em]{\shortstack[l]{\textbf{Additional}\\\textbf{modality}\\\textbf{gain cues}}}
& \texttt{router\_audio\_distracting\_content\_risk} & Risk that clips include distracting content \\
& \texttt{router\_graph\_target\_node\_coverage} & Target-entity coverage among graph nodes \\
& \texttt{router\_graph\_subgraph\_confidence} & Confidence of target-focused subgraph \\
& \texttt{router\_graph\_relation\_match} & Graph--question relation match \\
& \texttt{router\_graph\_irrelevant\_node\_ratio} & Fraction of irrelevant graph nodes \\
& \texttt{router\_graph\_spurious\_relation\_risk} & Risk of spurious graph relations \\
& \texttt{router\_graph\_clutter\_risk} & Risk of graph clutter or noise \\
& \texttt{router\_graph\_attribute\_loss\_risk} & Risk of attribute loss in graph abstraction \\
& \texttt{router\_graph\_marginal\_relation\_gain} & Relation-coverage gain from graph \\
& \texttt{router\_audio\_graph\_overlap} & Audio--graph evidence overlap \\
& \texttt{similarity\_q\_initial} & Cosine similarity: question vs.\ initial answer \\
& \texttt{similarity\_q\_critique} & Cosine similarity: question vs.\ self-critique \\
& \texttt{similarity\_q\_evidence} & Cosine similarity: question vs.\ evidence \\
& \texttt{similarity\_q\_description} & Cosine similarity: question vs.\ clip descriptions \\
& \texttt{similarity\_q\_all} & Cosine similarity: question vs.\ all text artifacts \\
 
\end{xltabular}
}

 
\clearpage
{\footnotesize
\setlength{\tabcolsep}{2.5pt}
\setlength{\LTpre}{0pt}
\setlength{\LTpost}{6pt}
 
\begin{xltabular}{\textwidth}{@{}p{1.8cm}lX@{}}

\caption{List of 100 Video-QA router features organized by routing-risk group.
The first three groups primarily detect \emph{under-fusion} (missing decisive evidence); the fourth group primarily reduces \emph{over-fusion} (unnecessary token cost).}
\label{tab:video_feature_definitions} \\

\toprule
\textbf{Group} & \textbf{Feature Name} & \textbf{Description} \\
\midrule
\endfirsthead
 
\toprule
\textbf{Group} & \textbf{Feature Name} & \textbf{Description} \\
\midrule
\endhead
 
\midrule
\multicolumn{3}{r}{\itshape Continued on next page} \\
\endfoot
\bottomrule
\\[-6pt]

\endlastfoot
 
& \texttt{target\_nouns} & Key nouns/entities from question and options \\
& \texttt{target\_nouns\_len} & \# of extracted target nouns \\
& \texttt{target\_nouns\_\{0,1,2,3\}} & Target noun at index 0--3 \textbf{(4 features)} \\
& \texttt{demand} & Structured question demand summary \\
& \texttt{demand\_question\_category} & Assigned high-level question categories \\
& \texttt{demand\_question\_category\_len} & \# of assigned categories \\
& \texttt{demand\_question\_category\_\{0,1\}} & Category at index 0, 1 \textbf{(2 features)} \\
& \texttt{demand\_target\_noun\_count} & \# of target nouns referenced in question \\
\multirow{-9}{*}[-4em]{\shortstack[l]{\textbf{Evidence}\\\textbf{requirement}\\\textbf{cues}}}
  & \texttt{demand\_requires\_action\_recognition} & Requires understanding actions or activities \\
& \texttt{demand\_requires\_temporal\_reasoning} & Requires reasoning about event sequence \\
& \texttt{demand\_requires\_spatial\_reasoning} & Requires spatial reasoning within frames \\
& \texttt{demand\_requires\_appearance\_recognition} & Requires recognizing visual appearance details \\
& \texttt{demand\_requires\_ocr} & Requires reading in-frame text (OCR) \\
& \texttt{demand\_requires\_counting} & Requires counting entities across frames \\
& \texttt{demand\_relation\_cue\_count} & \# of relation/temporal cue terms in question \\
& \texttt{text\_evidence\_to\_confirm} & Video evidence needed to confirm hypothesis \\
 
\midrule
 
& \texttt{text} & Structured text-only output \\
& \texttt{text\_initial\_answer} & Short initial answer justification from text output \\
& \texttt{text\_confidence} & Text output confidence score [0--1] \\
& \texttt{text\_reasoning} & Short reasoning supporting the initial answer \\
& \texttt{text\_signals} & Diagnostic signals from text artifact \\
& \texttt{text\_signals\_has\_hedges} & Whether hedge terms (e.g., \emph{maybe}) appear \\
& \texttt{text\_signals\_hedge\_terms\_found} & List of detected hedge terms \\
& \texttt{text\_signals\_cites\_visual\_evidence} & Whether modality-specific evidence is cited \\
\multirow{-9}{*}[-4em]{\shortstack[l]{\textbf{Uncertainty}\\\textbf{cues}}}
  & \texttt{text\_signals\_generic\_priors} & Whether generic priors (e.g., \emph{usually}) are used \\
& \texttt{text\_signals\_prior\_terms\_found} & List of detected generic-prior terms \\
& \texttt{text\_signals\_prior\_terms\_len} & \# of generic-prior terms found \\
& \texttt{text\_signals\_prior\_terms\_0} & First generic-prior term found \\
& \texttt{text\_self\_critique} & Self-critique of the hypothesis answer \\
& \texttt{text\_self\_critique\_neg\_reasoning} & Counter-argument against the hypothesis \\
& \texttt{router\_text\_confidence} & Router's text-artifact confidence score \\
& \texttt{router\_confidence\_calibration\_flag} & Calibration: overconfident / underconfident / ok \\
 
\midrule
 
& \texttt{video} & Structured video-grounded output \\
& \texttt{video\_clips} & List of proposed video region clips \\
& \texttt{video\_clip\_$k$} & Record for clip $k$, $k\!\in\!\{0,1,2\}$ \textbf{(3 features)} \\
& \texttt{video\_clip\_$k$\_start} & Start timestamp (seconds) for clip $k$ \textbf{(3 features)} \\
& \texttt{video\_clip\_$k$\_end} & End timestamp (seconds) for clip $k$ \textbf{(3 features)} \\
& \texttt{video\_clip\_$k$\_confidence} & Confidence clip $k$ has decisive evidence \textbf{(3 features)} \\
& \texttt{video\_clip\_$k$\_description} & Short description of clip $k$ content \textbf{(3 features)} \\
& \texttt{video\_clip\_$k$\_description\_relevance} & Description relevance to question for clip $k$ \textbf{(3 features)} \\
& \texttt{video\_clip\_$k$\_contains\_targets} & Whether clip $k$ description mentions targets \textbf{(3 features)} \\
& \texttt{video\_clip\_$k$\_matched\_nouns} & Target nouns matched in clip $k$ \textbf{(3 features)} \\
& \texttt{video\_clip\_\{0,1\}\_matched\_nouns\_len} & \# of matched target nouns in clip 0, 1 \textbf{(2 features)} \\
& \texttt{video\_clip\_\{0,1\}\_matched\_nouns\_0} & First matched noun for clip 0, 1 \textbf{(2 features)} \\
\multirow{-13}{*}[-7em]{\shortstack[l]{\textbf{Video}\\\textbf{evidence}\\\textbf{availability}\\\textbf{cues}}}
  & \texttt{video\_clip\_0\_matched\_nouns\_1} & Second matched noun for clip 0 \\
& \texttt{router\_best\_description\_relevance} & Max description relevance across all clips \\
& \texttt{router\_clip\_relevance\_mean} & Mean description relevance across clips \\
& \texttt{router\_clip\_relevance\_std} & Std.\ of description relevance across clips \\
& \texttt{router\_clip\_relevance\_min} & Min description relevance across clips \\
& \texttt{router\_clip\_relevance\_max} & Max description relevance across clips \\
& \texttt{router\_low\_relevance\_clip\_ratio} & Fraction of clips with low relevance \\
& \texttt{router\_clip\_confidence\_variance} & Variance of clip confidence scores \\
& \texttt{router\_clip\_relevance\_distribution} & Bucketed clip relevance distribution \\
& \texttt{router\_clip\_rel\_dist\_very\_low} & Share of clips in very-low relevance bucket \\
& \texttt{router\_clip\_rel\_dist\_low} & Share of clips in low relevance bucket \\
& \texttt{router\_clip\_rel\_dist\_medium} & Share of clips in medium relevance bucket \\
& \texttt{router\_clip\_rel\_dist\_high} & Share of clips in high relevance bucket \\
 
\midrule
 
& \texttt{router} & Structured router feature bundle \\
& \texttt{router\_video\_marginal\_coverage\_gain} & Estimated target-coverage gain from video \\
& \texttt{router\_text\_video\_overlap} & Text--video evidence overlap \\
& \texttt{router\_video\_irrelevant\_entity\_risk} & Risk that clips emphasize irrelevant entities \\
\multirow{-5}{*}[-2em]{\shortstack[l]{\textbf{Additional}\\\textbf{modality}\\\textbf{gain cues}}}
& \texttt{router\_video\_distracting\_context\_risk} & Risk that clips include distracting context \\
& \texttt{router\_graph\_target\_node\_coverage} & Target-entity coverage among graph nodes \\
& \texttt{router\_graph\_subgraph\_confidence} & Confidence of target-focused subgraph \\
& \texttt{router\_graph\_relation\_match} & Graph--question relation match \\
& \texttt{router\_graph\_irrelevant\_node\_ratio} & Fraction of irrelevant graph nodes \\
& \texttt{router\_graph\_spurious\_relation\_risk} & Risk of spurious graph relations \\
& \texttt{router\_graph\_clutter\_risk} & Risk of graph clutter or noise \\
& \texttt{router\_graph\_attribute\_loss\_risk} & Risk of attribute loss in graph abstraction \\
& \texttt{router\_graph\_marginal\_relation\_gain} & Relation-coverage gain from graph \\
& \texttt{router\_video\_graph\_overlap} & Video--graph evidence overlap \\
& \texttt{similarity\_q\_initial} & Cosine similarity: question vs.\ initial answer \\
& \texttt{similarity\_q\_critique} & Cosine similarity: question vs.\ self-critique \\
& \texttt{similarity\_q\_evidence} & Cosine similarity: question vs.\ evidence \\
& \texttt{similarity\_q\_description} & Cosine similarity: question vs.\ clip descriptions \\
& \texttt{similarity\_q\_all} & Cosine similarity: question vs.\ all text artifacts \\
& \texttt{similarity\_video\_clipmean} & Cosine similarity: video vs.\ mean clip embedding \\
& \texttt{similarity\_video\_clipmax} & Cosine similarity: video vs.\ max clip embedding \\
 
\end{xltabular}
}
 
 
\section{Best Combination Distribution for Audio-QA and Video-QA}
\label{sec:oracle-distribution-appendix}

We extend the best combination analysis from VQA (Section~\ref{sec:insight}) to Audio-QA and Video-QA. For each instance, we define the best combination as the correct one with the lowest token cost, or the lowest-cost combination if none is correct. We group questions into task-specific question types.

\begin{figure}[t]
    \centering
    \begin{subfigure}{0.49\textwidth}
        \centering
        \includegraphics[width=\textwidth]{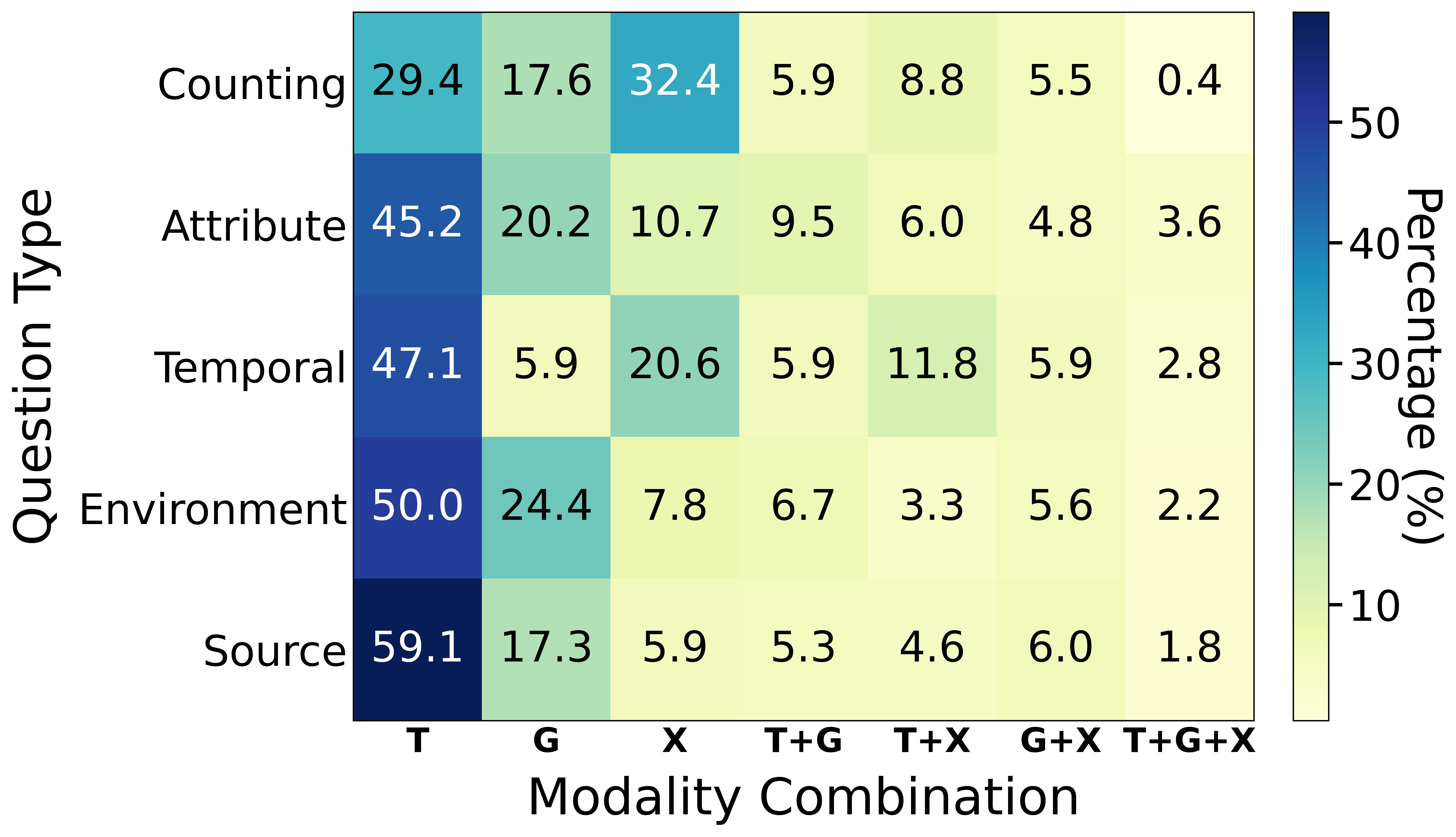}
        \caption{Audio-QA}
        \label{fig:audio_modality_heatmap}
    \end{subfigure}\hfill
    \begin{subfigure}{0.49\textwidth}
        \centering
        \includegraphics[width=\textwidth]{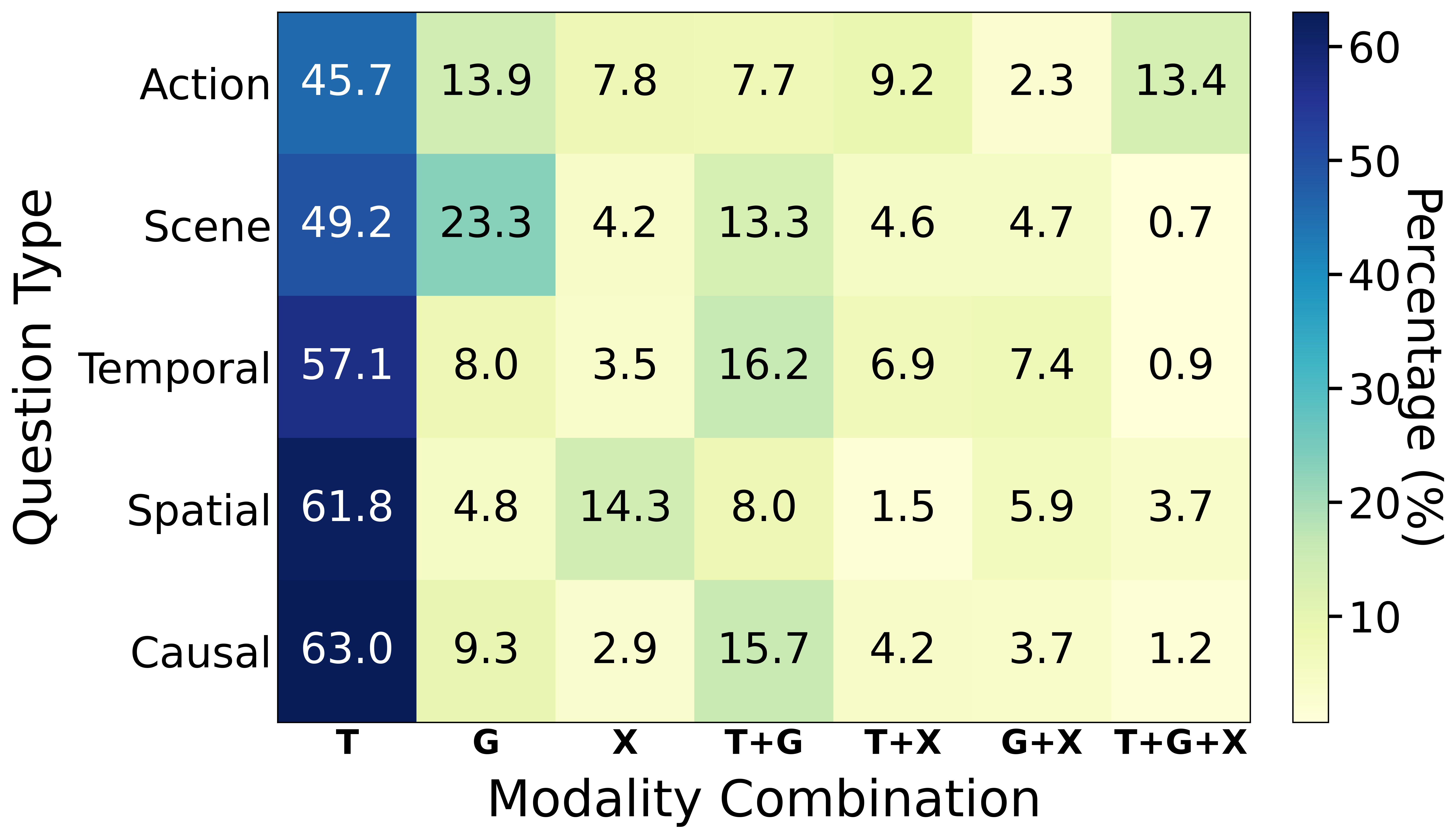}
        \caption{Video-QA}
        \label{fig:video_modality_heatmap}
    \end{subfigure}
    \caption{Best combination distribution across five question types. Each row sums to 100\%.}
    \label{fig:modality_heatmap_appendix}
\end{figure}

\paragraph{Best Combination Analysis for Audio-QA.}
We group Audio-QA instances into five question types: (\romannumeral1) \emph{counting}, which asks how many auditory events or entities are present; (\romannumeral2) \emph{attribute}, which asks about acoustic properties, tone, emotion, or musical characteristics; (\romannumeral3) \emph{temporal}, which asks about temporal ordering or duration of sounds; (\romannumeral4) \emph{environment}, which asks about scene or setting identification; and (\romannumeral5) \emph{source}, which asks what produces a sound. As shown in Fig.~\ref{fig:audio_modality_heatmap}, the best combination distribution varies across question types. 
Counting is the only question type where the task-dependent modality $\mathrm{X}$ (32.4\%) is higher than $\mathrm{T}$ (29.4\%), indicating that the raw audio signal is essential for distinguishing auditory event counts. For attribute and environment questions, graph-based interaction $\mathrm{G}$ is more frequently the best combination (20.2\% and 24.4\%) than other non-text modalities, reflecting that structured relational cues help ground fine-grained properties and scene context. For temporal questions, $\mathrm{X}$ (20.6\%) and $\mathrm{T+X}$ (11.8\%) benefit more from direct access to the audio signal, since reasoning about the order of auditory events requires hearing the original sequence. Source identification is the question type that prefers $\mathrm{T}$ most (59.1\%), yet over 40\% of instances still benefit from non-text modalities.

\paragraph{Best Combination Analysis for Video-QA.}
We group Video-QA instances into five question types: (\romannumeral1) \emph{action}, which asks about behaviors or activities; (\romannumeral2) \emph{scene}, which asks about visual setting, appearance, or emotional state; (\romannumeral3) \emph{temporal}, which asks about temporal ordering of events; (\romannumeral4) \emph{spatial}, which asks about spatial relationships or counting entities across frames; and (\romannumeral5) \emph{causal}, which asks about motivations or causes. As shown in Fig.~\ref{fig:video_modality_heatmap}, the best combination distribution again varies by question type but differs from VQA and Audio-QA. Action questions benefit from a wider range of combinations, with $\mathrm{T}$ at 45.7\%, $\mathrm{G}$ at 13.9\%, and $\mathrm{T+G+X}$ at 13.4\%, showing that recognizing behaviors and activities can require both structured relations and multimodal evidence. $\mathrm{T+G}$ is consistently strong across temporal (16.2\%) and causal (15.7\%) questions, suggesting that graph-structured relations help ground event sequences and causal chains. For spatial questions, $\mathrm{X}$ benefits more (14.3\%) than in other question types, since raw video frames provide direct spatial layout information. Scene questions show the strongest $\mathrm{G}$ contribution (23.3\%), reflecting that structured scene graphs effectively capture visual context. 

\paragraph{Cross-Task Comparison.}
Across all three task families, no single modality combination is best for every question type, consistent with the findings for VQA in Section~\ref{sec:insight}.  
The best combination varies by task. In VQA, $\mathrm{T+X}$ is most prominent (Fig.~\ref{fig:modality_heatmap}). In Audio-QA and Video-QA, $\mathrm{T}$ is more likely to be selected, but $\mathrm{X}$ and $\mathrm{G}$ benefit more in Audio-QA while $\mathrm{T+G}$ benefits more in Video-QA.
These differences reflect the token cost of each task-dependent modality. In VQA and Audio-QA, the task-dependent modality uses fewer tokens, so combinations that include it can achieve higher accuracy while keeping token cost low. In Video-QA, the task-dependent modality uses substantially more tokens, so the best combination more often favors $\mathrm{T+G}$, which provides structured evidence beyond text alone. These results reinforce the motivation for adaptive modality selection, since the best combination depends not only on the question type but also on the token cost of the task-dependent modality in each task family.

\section{Latency and Cost Analyses}
\label{sec:cost-analyses}

In this section, we analyze the latency and cost of \name. We evaluate the latency overhead of the routing pipeline in Appendix~\ref{sec:latency}, report the token cost of \name by phase in Appendix~\ref{sec:token-breakdown}, and quantify the one-time offline cost of training the routers in Appendix~\ref{sec:offline-cost}.

\subsection{Latency Overhead Analysis}
\label{sec:latency}

To evaluate the latency overhead of \name, we measure the end-to-end latency per instance of \name and of the seven fixed combinations on all eight benchmarks. The latency overhead of \name covers the structured baseline response, feature extraction, and adaptive modality routing, relative to directly executing MAD with all seven fixed-modality baselines. We report the average latency per instance in seconds in Table~\ref{tab:latency}. \name's feature extraction and routing add negligible overhead to the total latency (less than 0.3s per instance in our measurements). The remaining latency stems from the structured baseline LLM response and MAD execution, both required by any MAD method. On average, the end-to-end latency of \name is 16.5s for VQA, 11.7s for Audio-QA, and 15.4s for Video-QA. By adaptively selecting the best modality combination per instance, \name balances accuracy and latency. The total latency depends on multiple factors, including the number of modalities processed, the input size of each modality (e.g., uploading and encoding video is much slower than text), the prompt length during debate, and the LLM generation time. Since \name selects the modality combination that can answer correctly at low token cost, it avoids expensive modalities when they are unnecessary, which often reduces latency as well. Across the eight benchmarks, \name is faster than the highest-accuracy fixed-modality combination on four datasets while achieving higher accuracy on all eight. On OKVQA, \name is faster than $\mathrm{T+G+X}$ (14.0s vs. 21.3s) while improving accuracy by 0.3\%. On NextQA, \name is faster than $\mathrm{T+G+X}$ (14.6s vs. 28.2s) while improving accuracy by 1.6\%. On IntentQA, \name is faster than $\mathrm{T+X}$ (16.1s vs. 19.4s) while improving accuracy by 4.2\%. On VCR, \name is comparable to $\mathrm{G+X}$ (22.0s vs. 22.5s) while improving accuracy by 3.1\%. On the remaining four datasets (GQA, VQA-v2, Clotho, and DCASE), \name is 1.5-6.8s slower than the highest-accuracy fixed combination, but these latency increases are modest relative to the accuracy gains of 0.9\%-3.7\%.

\begin{table}[t]
\centering
\caption{Latency comparison across eight multimodal QA benchmarks (in seconds). Bold indicates the lowest latency for each dataset.}
\label{tab:latency}
\vspace{0.5em}
\begin{tabular}{lcccccccc}
\toprule
\multicolumn{1}{c}{\multirow{2.5}{*}{\textbf{Dataset}}}
& \multicolumn{3}{c}{\textbf{One Modality}}
& \multicolumn{3}{c}{\textbf{Two Modalities}}
& \multicolumn{1}{c}{\textbf{All}}
& \multicolumn{1}{c}{\textbf{Adaptive}} \\
\cmidrule(lr){2-4} \cmidrule(lr){5-7} \cmidrule(lr){8-8} \cmidrule(lr){9-9}
& $\mathrm{T}$ & $\mathrm{G}$ & $\mathrm{X}$ & $\mathrm{T+G}$ & $\mathrm{T+X}$ & $\mathrm{G+X}$ & $\mathrm{T+G+X}$ & \name \\
\midrule
GQA    & 8.4 & 15.3 & \textbf{6.8} & 8.2 & 10.0 & 16.7 & 18.1 & 16.4 \\
OKVQA  & 8.7 & 12.1 & \textbf{6.4} & 8.3 & 6.7 & 19.0 & 21.3 & 14.0 \\
VCR    & 12.5 & 14.7 & \textbf{8.8} & 14.2 & 15.7 & 22.5 & 25.4 & 22.0 \\
VQA-v2 & 8.0 & 10.7 & \textbf{6.1} & 8.2 & 6.8 & 18.2 & 21.4 & 13.6 \\ \midrule
Clotho    & 6.0 & \textbf{5.8} & 9.8 & 6.2 & 9.7 & 10.0 & 10.2 & 11.7 \\
DCASE & 5.3 & 9.1 & \textbf{5.2} & 6.9 & 9.1 & 10.7 & 17.3 & 11.6 \\ \midrule
IntentQA    & \textbf{9.8} & 15.6 & 28.7 & 10.2 & 19.4 & 22.3 & 20.2 & 16.1 \\
NextQA & \textbf{10.1} & 18.2 & 29.1 & 13.1 & 25.3 & 22.9 & 28.2 & 14.6 \\
\bottomrule
\end{tabular}
\end{table}

\subsection{Token Cost Analysis}
\label{sec:token-breakdown}

The token cost of \name in Table~\ref{tab:main-results} covers both the Phase~{\large \textcircled{\small 1}} structured baseline response and the final MAD execution with the selected modality combination, whereas feature extraction in Phase~{\large \textcircled{\small 2}} and the routing decision in Phase~{\large \textcircled{\small 3}} are local computations that do not consume LLM tokens and add less than 0.3s latency per instance (Appendix~\ref{sec:latency}). We report the two parts of the token cost of \name in Table~\ref{tab:token-breakdown}, averaged over the benchmarks of each task family, and make two observations. First, the MAD execution accounts for most of the token cost of \name, with 68.4\%, 66.2\%, and 83.7\% on VQA, Audio-QA, and Video-QA, respectively, and the share is largest on Video-QA since the video clips exchanged in the debate are expensive. Second, the structured baseline response is cheap relative to the debate it serves. For example, on Video-QA, the structured baseline response uses 7,683 tokens, whereas the MAD execution uses 39,511 tokens. Therefore, the token cost of \name is determined by the modality combination selected for the debate rather than by the structured baseline response.

\begin{table}[ht]
\centering
\caption{Average token cost per instance of \name, split into the Phase~{\large \textcircled{\small 1}} structured baseline agent and the final MAD execution with the selected modality combination.}
\label{tab:token-breakdown}
\vspace{0.5em}
\begin{tabular}{lccc}
\toprule
\textbf{Task Family} & \textbf{Structured Baseline Response} & \textbf{MAD Execution} & \textbf{Total} \\
\midrule
VQA      & 2,203 & 4,773  & 6,976 \\
Audio-QA & 1,298 & 2,545  & 3,843 \\
Video-QA & 7,683 & 39,511 & 47,194 \\
\bottomrule
\end{tabular}
\end{table}

\subsection{Offline Training Cost and Amortization}
\label{sec:offline-cost}

To quantify the one-time offline cost of \name, we report the training cost of the adaptive modality router for each task family, amortize the training cost over inference queries.

\paragraph{Training Cost.}
Training the adaptive modality router requires a one-time data collection. In the data collection, we run MAD under each of the seven modality combinations on every auxiliary instance and record the correctness and token cost of each modality combination. Each auxiliary instance therefore requires one debate per candidate modality combination, and the number of candidate modality combinations is $2^M-1$ for $M$ interaction modalities. \name uses $M=3$, and a new task family changes what the task-dependent modality $\mathrm{X}$ represents without increasing $M$. Moreover, the data collection is not repeated at inference, since MAD executes only the selected modality combination. We report the data collection cost and the router training time for each task family in Table~\ref{tab:offline-cost} and make two observations. First, the data collection uses 172.64M tokens in total at an estimated API cost of 195.25 USD, and Video-QA accounts for most of the cost with 137.35M tokens, since video inputs are expensive. Second, training the three routers takes only 47.9s in total on one NVIDIA RTX 4090 GPU. Therefore, the one-time cost of \name stems almost entirely from running MAD on the auxiliary datasets, and this cost is paid once per task family.

\begin{table}[ht]
\centering
\caption{Training cost of the adaptive modality router for each task family.}
\label{tab:offline-cost}
\vspace{0.5em}
\begin{tabular}{lcccc}
\toprule
\multirow{2}{*}{\textbf{Task Family}} & \textbf{Auxiliary} & \textbf{Collection} & \textbf{Estimated API} & \textbf{Router Training} \\
& \textbf{Instances} & \textbf{Tokens} & \textbf{Cost (USD)} & \textbf{Time (s)} \\
\midrule
VQA      & 500   & 27.73M  & 31.36  & 19.3 \\
Audio-QA & 250   & 7.56M   & 8.55   & 13.4 \\
Video-QA & 250   & 137.35M & 155.34 & 15.2 \\
\midrule
Total    & 1,000 & 172.64M & 195.25 & 47.9 \\
\bottomrule
\end{tabular}
\end{table}

\paragraph{Amortization.}
To evaluate how fast the one-time cost is recovered, we amortize the one-time cost over the evaluation queries in our study. The amortized cost is 107 tokens and 0.00012 USD per VQA query, 1,443 tokens and 0.0016 USD per Audio-QA query, and 19,337 tokens and 0.022 USD per Video-QA query. The amortized cost can be further reduced as each router serves more queries, since each router is trained once on the auxiliary datasets and then applied to every benchmark of the same task family without retraining, as in our evaluation. Video-QA has the largest one-time cost, but Video-QA also has the largest saving at inference. On NextQA, \name achieves 78.8\% accuracy with 48,636 tokens per query, whereas the most accurate fixed combination $\mathrm{T+G+X}$ achieves 77.2\% accuracy with 99,765 tokens per query. The saving is 51,129 tokens per query, and therefore the saving recovers the full Video-QA data collection cost of 137.35M tokens after about 2,686 queries.

\section{Ablation Studies}
\label{sec:ablation-studies}

In this section, we present two ablation studies on the design of \name. We compare the structured routing prompt with a standard CoT prompt in Appendix~\ref{sec:ablation-cot}, and evaluate the contribution of each feature group by removing one feature group at a time in Appendix~\ref{subsec:logo}.

\subsection{Structured Routing Prompt vs. CoT Prompt}
\label{sec:ablation-cot}

A key design choice in \name is the structured modality-routing prompt (Phase~{\large \textcircled{\small 1}}), which generates not only an initial answer but also explicit confidence scores, self-critique, missing-evidence statements, and task-dependent modality summaries. These structured outputs enable the extraction of a set of (109 for VQA, 98 for Audio-QA, 100 for Video-QA)  interpretable routing features that drive the adaptive modality router. To evaluate the importance of this design, we replace the structured prompt with a standard CoT prompt that produces only a reasoning trace and a final answer, and retrain the router using only features extractable from CoT outputs (e.g., answer length, token count, text embeddings of the reasoning trace).
All other components, including the router classifiers, token-cost regressors, and the routing decision rule, remain identical.

As shown in Table~\ref{tab:ablation-cot}, replacing the structured routing prompt with a standard CoT prompt consistently degrades accuracy across all eight benchmarks. The structured prompt improves accuracy by 3.0\% on GQA, 5.0\% on VQA-v2, 5.5\% on Clotho, and 10.2\% on IntentQA, with an average gain of 4.45\% across all benchmarks. These improvements are especially large on audio and video tasks, where the CoT-based router tends to favor cheaper (i.e., lower token cost) text-based modality since audio-based and video-based evidence is more token-expensive and its usefulness is not explicitly represented in the CoT output. Although the CoT variant can still extract weak routing cues from the reasoning trace, the structured prompt provides more reliable confidence, missing-evidence, and modality-usefulness signals for adaptive routing.

The token-cost results further reveal that the structured prompt improves routing quality in two complementary ways. On GQA, OKVQA, VCR, and DCASE, it achieves both higher accuracy and lower token cost. For example, compared with the CoT prompt, the structured prompt reduces the average token cost by 300 tokens on GQA, 380 tokens on OKVQA, 679 tokens on VCR, and 151 tokens on DCASE, while still improving accuracy on all four datasets. This indicates that the structured routing features help the router avoid over-fusion: when additional modalities are unlikely to help, the router can select the modality combination with lower token costs without sacrificing accuracy. However, on VQA-v2, Clotho, IntentQA, and NextQA, the structured prompt uses more tokens but obtains substantially higher accuracy. The increase of token usage is small on VQA-v2 and Clotho, where the structured prompt uses only 102 and 80 additional tokens, respectively, but yields sizable accuracy gains of 5.0\% and 5.5\%. The effect is more pronounced on the video benchmarks: the structured prompt uses 744 additional tokens on IntentQA and 7,759 additional tokens on NextQA, but improves accuracy by 10.2\% and 3.4\%. These results suggest that the CoT-based router often fails to incorporate sufficient task-dependent evidence, especially on audio and video benchmarks where the relevant modality is more expensive to process. As the CoT prompt does not explicitly indicate whether the relevant audio segments or spatio-temporal video clips are available, informative, or necessary, the router tends to favor cheaper text-dominant choices. Although this behavior can reduce token cost in some cases, it also causes the model to miss critical decisive evidence.

Importantly, the CoT-based router is not entirely ineffective. Although CoT outputs provide only weak routing cues, such as reasoning length, token count, and semantic information in the generated rationale, the adaptive modality selection mechanism can still extract some useful features from them and may therefore outperform a text-only interaction strategy. However, the consistent accuracy gap between the CoT variant and the structured modality-routing prompt demonstrates that these implicit signals alone are insufficient for accurate modality routing. The structured prompt provides explicit signals about confidence, self-critique, missing evidence, and task-dependent modality usefulness, enabling the router to distinguish cases in which a lower-cost modality combination is sufficient from those requiring richer multimodal evidence. These results demonstrate that the structured modality-routing prompt is essential for adaptive modality routing.

\begin{table}[h]
\centering
\caption{Ablation study on routing prompt design for \name across eight benchmarks. We compare \name with a standard CoT prompt and the full structured routing prompt.}
\vspace{0.5em}
\label{tab:ablation-cot}{%
\begin{tabular}{lcccc}
\toprule
\multirow{2}{*}{\textbf{Dataset}}
& \multicolumn{2}{c}{\textbf{CoT Prompt}}
& \multicolumn{2}{c}{\textbf{Structured Routing Prompt}} \\
\cmidrule(lr){2-3} \cmidrule(lr){4-5}
& \textbf{Acc. (\%)} & \textbf{\# Tokens}
& \textbf{Acc. (\%)} & \textbf{\# Tokens} \\
\midrule
GQA
& 77.1 & 6,634
& \textbf{80.1} & \textbf{6,334} \\

OKVQA
& 91.1 & 7,155
& \textbf{93.8} & \textbf{6,775} \\

VCR
& 81.0 & 8,467
& \textbf{84.2} & \textbf{7,788} \\

VQA-v2
& 82.7 & \textbf{6,904}
& \textbf{87.7} & 7,006 \\

\midrule

Clotho
& 41.3 & \textbf{3,715}
& \textbf{46.8} & 3,795 \\

DCASE
& 49.2 & 4,042
& \textbf{51.8} & \textbf{3,891} \\
\midrule

IntentQA
& 65.2 & \textbf{45,008}
& \textbf{75.4} & 45,752 \\

NextQA
& 75.4 & \textbf{40,877}
& \textbf{78.8} & 48,636 \\
\bottomrule
\end{tabular}%
}
\end{table}

\subsection{Leave-One-Feature-Group-Out Ablation}
\label{subsec:logo}

To evaluate the contribution of each feature group beyond the SHAP analysis in Appendix~\ref{subsec:shap-results}, we remove one of the four feature groups from the routing features, retrain the correctness classifier and the token-cost regressor on the remaining three feature groups with the same auxiliary datasets, and evaluate the retrained router on all eight benchmarks. We report the accuracy and token cost of each retrained router in Table~\ref{tab:logo}, averaged within each task family. We find that removing any one feature group lowers accuracy and increases token cost in every task family, so all four feature groups contribute to the routing decision. The most important feature group differs across task families. Removing the uncertainty cues causes the largest accuracy drop on VQA, at 0.9\%, whereas removing the task-dependent evidence availability cues causes the largest accuracy drop on Audio-QA and Video-QA, at 0.7\% and 1.5\%. In every task family, the two feature groups with the largest accuracy drops are also the two feature groups with the highest SHAP importance for the correctness classifier in Appendix~\ref{subsec:shap-results}. Moreover, removing any one feature group lowers accuracy by at most 1.5\%, which shows that the router does not depend on one feature group alone. Therefore, the four feature groups are complementary, and each feature group adds information that the other three feature groups do not provide.

\begin{table}[ht]
\centering
\caption{Accuracy (Acc) and average token cost (Tkn) of \name after removing one feature group and retraining the router, averaged within each task family. Accuracy is reported as a percentage (\%), and Tkn denotes the average per-instance token cost, where K denotes thousands.}
\label{tab:logo}
\vspace{0.5em}
\begin{tabular}{lcccccc}
\toprule
\multicolumn{1}{c}{\multirow{2.5}{*}{\textbf{Removed Feature Group}}}
& \multicolumn{2}{c}{\textbf{VQA}}
& \multicolumn{2}{c}{\textbf{Audio-QA}}
& \multicolumn{2}{c}{\textbf{Video-QA}} \\
\cmidrule(lr){2-3} \cmidrule(lr){4-5} \cmidrule(lr){6-7}
& \textbf{Acc} & \textbf{Tkn} & \textbf{Acc} & \textbf{Tkn} & \textbf{Acc} & \textbf{Tkn} \\
\midrule
None (\name)                                & 86.5 & 6,976 & 49.3 & 3,843 & 77.1 & 47.2K \\
\midrule
Evidence requirement cues                   & 86.0 & 7,596 & 48.9 & 4,183 & 76.7 & 49.5K \\
Uncertainty cues                            & 85.6 & 7,432 & 48.9 & 4,097 & 76.8 & 48.1K \\
Task-dependent evidence availability cues   & 85.8 & 7,749 & 48.6 & 4,142 & 75.6 & 52.4K \\
Additional modality gain cues               & 86.1 & 7,081 & 48.7 & 3,994 & 75.8 & 49.8K \\
\bottomrule
\end{tabular}
\end{table}

\section{Cross-Backbone Generalization: GPT-5.4-nano}
\label{sec:gpt-backbone}

To verify whether \name generalizes across LLM backbones, we replace Gemini 2.5 Flash with GPT-5.4-nano. As GPT-5.4-nano does not support audio or video inputs, we restrict this study to the four VQA benchmarks. All other components, including the graph pipeline, crop tool, debate protocol, router configuration, and feature extraction, are kept identical to the Gemini 2.5 Flash experiments.

\paragraph{Results.}
As shown in Table~\ref{tab:gpt-backbone-results}, \name achieves the highest accuracy on all four VQA benchmarks under the GPT-5.4-nano backbone while remaining token-efficient. Compared with the highest-accuracy fixed-modality baseline on each dataset, \name improves accuracy by 7.1\% on GQA, 8.8\% on OKVQA, 10.4\% on VCR, and 9.3\% on VQA-v2, with an average accuracy gain of 8.9\%. \name also reduces token cost by 9.4\%-55.3\% relative to the corresponding highest-accuracy fixed-modality baselines. These results show that adaptive interaction modality selection remains effective after changing the LLM backbone.

\paragraph{Comparison with Gemini 2.5 Flash.}
Compared with the Gemini 2.5 Flash results in Section~\ref{sec:evaluation}, \name's average accuracy gain over the best fixed-modality baseline increases from 2.4\% to 8.9\%. Under GPT-5.4-nano, all fixed-modality baselines achieve lower accuracy (e.g., $\mathrm{X}$ drops from 71.2\% to 60.8\% on GQA, $\mathrm{T+X}$ drops from 84.0\% to 64.9\% on VQA-v2). \name's absolute accuracy is also lower than under Gemini 2.5 Flash (e.g., 74.4\% vs. 80.1\% on GQA). However, adaptive modality selection yields a much larger accuracy gain with GPT-5.4-nano. This suggests that \name is particularly useful when fixed-modality baselines have lower accuracy, because the router can recover additional correct answers by choosing different modality combinations for different instances. Across both backbones, \name consistently outperforms every fixed-modality baseline, showing that the routing mechanism generalizes across LLM backbones with different reasoning capabilities.

\paragraph{Why \name Improves More on GPT-5.4-nano.}
The larger accuracy gain on GPT-5.4-nano arises since different modality combinations on GPT-5.4-nano tend to answer different types of instances correctly. Even when each fixed combination individually achieves low accuracy, different combinations still answer different subsets of instances correctly. Thus, the set of instances answered correctly by at least one combination can be much larger than the set answered correctly by any single fixed combination. 
\name exploits this complementarity by selecting the best modality combination for each instance. Under GPT-5.4-nano, each fixed-combination baseline has lower accuracy individually, but their successes remain complementary: many instances are still answered correctly by at least one combination. This gives \name more room to improve accuracy through adaptive selection.

\begin{table*}[t]
\centering
\caption{Accuracy ({Acc}) and average token cost ({Tkn}) across four VQA benchmarks using the GPT-5.4-nano backbone. Accuracy is reported as a percentage (\%), and Tkn denotes the average per-instance token cost.}
\label{tab:gpt-backbone-results}

\resizebox{\textwidth}{!}
{%
\setlength{\tabcolsep}{2pt}
\begin{tabular}{lccccccccccccccccccccc}
\toprule
\multicolumn{1}{c}{\multirow{4}{*}{\textbf{Dataset}}}
& \multicolumn{6}{c}{\textbf{One Modality}}
& \multicolumn{6}{c}{\textbf{Two Modalities}}
& \multicolumn{2}{c}{\textbf{All}}
& \multicolumn{2}{c}{\textbf{Adaptive}} \\
\cmidrule(lr){2-7} \cmidrule(lr){8-13} \cmidrule(lr){14-15} \cmidrule(lr){16-17}
& \multicolumn{2}{c}{$\mathrm{T}$}
& \multicolumn{2}{c}{$\mathrm{G}$}
& \multicolumn{2}{c}{$\mathrm{X}$}
& \multicolumn{2}{c}{$\mathrm{T+G}$}
& \multicolumn{2}{c}{$\mathrm{T+X}$}
& \multicolumn{2}{c}{$\mathrm{G+X}$}
& \multicolumn{2}{c}{$\mathrm{T+G+X}$}
& \multicolumn{2}{c}{\name} \\
\cmidrule(lr){2-3}
\cmidrule(lr){4-5}
\cmidrule(lr){6-7}
\cmidrule(lr){8-9}
\cmidrule(lr){10-11}
\cmidrule(lr){12-13}
\cmidrule(lr){14-15}
\cmidrule(lr){16-17}
& \textbf{Acc} & \textbf{Tkn}
& \textbf{Acc} & \textbf{Tkn}
& \textbf{Acc} & \textbf{Tkn}
& \textbf{Acc} & \textbf{Tkn}
& \textbf{Acc} & \textbf{Tkn}
& \textbf{Acc} & \textbf{Tkn}
& \textbf{Acc} & \textbf{Tkn}
& \textbf{Acc} & \textbf{Tkn} \\
\midrule

GQA
& 58.8 & 2,093
& 59.0 & 2,097
& 60.8 & \textbf{1,854}
& 61.9 & 2,957
& 67.3 & 2,472
& 61.8 & 4,958
& 64.2 & 5,714
& \textbf{74.4} & 1,912 \\

OKVQA
& 68.6 &3,224
& 74.5 &3,191
& 75.5 &\textbf{2,525}
& 74.9 &3,862
& 78.4 &2,987
& 72.5 &5,161
& 76.8 &6,714
& \textbf{87.2} & 2,707\\

VCR
& 43.1 &10,277
& 45.1 &10,462
& 47.1 &9,521
& 47.6 &11,319
& 47.1 &9,179
& 49.0 &12,741
& 52.8 &19,515
& \textbf{63.2} & \textbf{8,723}\\

VQA-v2
& 51.0 &2,864
& 56.9 &2,942
& 62.7 &\textbf{2,741}
& 58.8 &3,786
& 64.9 &3,379
& 64.1 &5,029
& 62.0 &6,675
& \textbf{74.2} &2,832 \\
\bottomrule

\end{tabular}%
}

\end{table*}

\section{Robustness Analysis}
\label{sec:robustness}

To evaluate the robustness of \name to an unreliable structured baseline output, we modify the structured baseline output of Phase~{\large \textcircled{\small 1}} at evaluation time and measure the accuracy and token cost of \name on all eight benchmarks. The structured baseline output has three components, which are the initial answer with the confidence score, the self-critique with the missing-evidence statement, and the task-dependent modality summaries. We consider four modifications. In Appendix~\ref{subsec:robust-perturb}, we mask the key evidence words in the missing-evidence statement, or shift the confidence score of the initial answer and the coordinates, confidence scores, and relevance scores of the task-dependent modality summaries. In Appendix~\ref{subsec:robust-semantic}, we make the baseline agent fully confident in an incorrect initial answer by forcing the confidence score to its maximum, or move the regions, segments, or clips in the task-dependent modality summaries to wrong locations. In all cases, we use the router trained on the unmodified auxiliary datasets without retraining, so the results measure how the trained router responds to an unreliable input. We report the results in Table~\ref{tab:robustness}, where the last column is the result of \name without modification from Table~\ref{tab:main-results}.

\begin{table}[ht]
\centering
\caption{Accuracy ({Acc}) and average token cost ({Tkn}) of \name when one component of the structured baseline output is masked, shifted, or made incorrect at evaluation time, without retraining the router. The last column is the result of \name without modification. Accuracy is reported as a percentage (\%), and {Tkn} denotes the average per-instance token cost, where K denotes thousands.}
\label{tab:robustness}
\vspace{0.5em}
\resizebox{\textwidth}{!}
{%
\setlength{\tabcolsep}{4pt}
\begin{tabular}{lcccccccccc}
\toprule
\multicolumn{1}{c}{\multirow{3.5}{*}{\textbf{Dataset}}}
& \multicolumn{4}{c}{\textbf{Incomplete or Perturbed Output}}
& \multicolumn{4}{c}{\textbf{Incorrect Output}}
& \multicolumn{2}{c}{\multirow{2.5}{*}{\textbf{No Modification}}} \\
\cmidrule(lr){2-5} \cmidrule(lr){6-9}
& \multicolumn{2}{c}{\textbf{Masked Statement}}
& \multicolumn{2}{c}{\textbf{Shifted Scores}}
& \multicolumn{2}{c}{\textbf{Wrong Confidence}}
& \multicolumn{2}{c}{\textbf{Mislocated Regions}}
& \multicolumn{2}{c}{} \\
\cmidrule(lr){2-3} \cmidrule(lr){4-5} \cmidrule(lr){6-7} \cmidrule(lr){8-9} \cmidrule(lr){10-11}
& \textbf{Acc} & \textbf{Tkn}
& \textbf{Acc} & \textbf{Tkn}
& \textbf{Acc} & \textbf{Tkn}
& \textbf{Acc} & \textbf{Tkn}
& \textbf{Acc} & \textbf{Tkn} \\
\midrule
GQA      & 80.1 & 6,392 & 80.1 & 6,383 & 79.6 & 6,670 & 79.4 & 6,550 & 80.1 & 6,334 \\
OKVQA    & 93.6 & 6,734 & 93.7 & 6,855 & 93.6 & 6,881 & 93.5 & 6,983 & 93.8 & 6,775 \\
VCR      & 84.1 & 7,625 & 84.1 & 8,207 & 83.8 & 7,746 & 83.8 & 8,152 & 84.2 & 7,788 \\
VQA-v2   & 87.7 & 6,991 & 87.7 & 7,468 & 87.2 & 7,358 & 86.8 & 7,704 & 87.7 & 7,006 \\
\midrule
Clotho   & 46.3 & 3,817 & 46.7 & 3,877 & 46.5 & 3,760 & 46.2 & 4,018 & 46.8 & 3,795 \\
DCASE    & 51.7 & 3,841 & 49.6 & 3,949 & 51.4 & 3,794 & 50.9 & 4,492 & 51.8 & 3,891 \\
\midrule
IntentQA & 75.0 & 46.0K & 74.3 & 51.6K & 75.3 & 45.5K & 74.6 & 51.9K & 75.4 & 45.8K \\
NextQA   & 78.8 & 49.0K & 78.8 & 48.4K & 78.7 & 48.5K & 78.0 & 54.4K & 78.8 & 48.6K \\
\bottomrule
\end{tabular}%
}
\end{table}

\subsection{Incomplete or Perturbed Outputs}
\label{subsec:robust-perturb}

To evaluate \name when the structured baseline output is incomplete, we mask the key evidence words in the missing-evidence statement, such as the nouns that name the needed evidence, so the router no longer knows which evidence the baseline agent considers missing. We find that the masked statement has little effect on \name. The accuracy drops by only 0.16\% on average, and the token cost changes by at most 2.1\% on any benchmark, since the missing-evidence statement is only one of the evidence requirement cues and the router still receives the question type and the other three feature groups. To evaluate \name when the structured baseline output is perturbed, we add an independent Gaussian perturbation $\epsilon \sim \mathcal{N}(0, 0.1^2)$ to the confidence score of the initial answer and to every normalized numerical field of the task-dependent modality summaries, which are the bounding box coordinates for VQA, the start and end timestamps for Audio-QA and Video-QA, the confidence score of each summary, and the relevance score of each caption. As a result, the uncertainty cues and the evidence availability cues are extracted from perturbed values. We find that the shifted scores have a larger but still small effect. The accuracy drops by 0.45\% and the token cost increases by 3.7\% on average, and six of the eight benchmarks lose at most 0.1\% accuracy. The two exceptions are DCASE, where the accuracy drops by 2.2\%, and IntentQA, where the token cost increases by 12.7\%. This difference stems from the feature groups that each router relies on. According to the SHAP analysis in Appendix~\ref{subsec:shap-results}, the routers of Audio-QA and Video-QA rely most on the evidence availability cues and the additional modality gain cues, and both feature groups are extracted from the shifted scores of the summaries, so the shifted scores change more routing decisions in these two task families. Nevertheless, under both modifications, \name remains more accurate than the most accurate fixed combination on average in every task family. Therefore, an incomplete or perturbed structured baseline output changes the accuracy of \name by less than 0.5\% on average.

\subsection{Incorrect Outputs}
\label{subsec:robust-semantic}
To evaluate \name when the baseline agent is confident in a wrong answer, we set the confidence score to its maximum of 1.0 for every instance whose initial answer is incorrect and keep the other uncertainty cues unchanged. If the router trusted the confidence score alone, the router would select the text-only combination for these instances and answer incorrectly. We find that the wrong confidence has little effect on \name. The accuracy drops by only 0.31\% and the token cost increases by only 0.9\% on average, since the correctness classifier also uses the hedge terms, the self-critique, and the other three feature groups, which still identify the instances that need additional evidence. To evaluate \name when the task-dependent modality summaries point to the wrong part of the input, we randomly move one third of the regions, segments, or clips to locations that do not overlap the original locations in space or time, and extract the evidence availability cues again from the moved locations. We find that the mislocated regions have the largest effect among the four modifications, but \name remains accurate. The accuracy drops by 0.68\% on average, whereas the token cost increases by 8.5\% on average, and the token increase is larger on Audio-QA and Video-QA, at 10.7\% and 12.6\%, than on VQA, at 5.3\%. This result follows from the accuracy-first objective of \name, which minimizes token cost only among the combinations expected to answer correctly. When the evidence availability cues become unreliable, \name keeps accuracy by spending more tokens, and the increase is larger on Audio-QA and Video-QA, whose routers rely more on the evidence availability cues according to the SHAP analysis in Appendix~\ref{subsec:shap-results}. Nevertheless, under both modifications, \name remains at least as accurate as the most accurate fixed combination on all eight benchmarks. Therefore, \name remains accurate even when the baseline agent is confident in a wrong answer or points to the wrong part of the input.

\section{Comparison with Single-group Adaptive Routers}
\label{sec:simple-routing}

Section~\ref{sec:routing-analysis} compares \name with four single-group adaptive routers, each of which selects the modality combination with the routing features of a single feature group. In this section, we first list the exact routing features that each single-group adaptive router uses in Appendix~\ref{subsec:simple-details}. We then compare the robustness of \name and the four single-group adaptive routers in Appendix~\ref{subsec:robust-simple}, where we flip, mask, or shift the routing features of each single-group adaptive router and measure how much the accuracy of each method drops.

\subsection{Details of the Single-group Adaptive Routers}
\label{subsec:simple-details}

Each single-group adaptive router selects the modality combination with the routing features of a single feature group, whereas the adaptive modality router of \name uses all four feature groups. Question-type routing uses only the question type features, which are the evidence requirement cues without the missing-evidence statement and include the question categories, the referenced entities, and the question demand flags, such as whether the question requires counting or spatial reasoning. Similarly, missing-evidence-only routing uses only the missing-evidence statement, which states what additional evidence would confirm or refute the initial answer, and the task-dependent modality evidence availability cues, which estimate whether the missing evidence is present in the input. Moreover, confidence-only routing uses only the uncertainty cues, which include the confidence score, hedge terms, generic priors, and the self-critique. Finally, the cost-aware heuristic uses only the additional modality gain cues, which estimate how much new evidence each modality adds and the risk of adding it. We give the exact feature names used by each single-group adaptive router for each task family in Table~\ref{tab:simple-features}, following the feature tables in Appendix~\ref{subsec:feature-design}.

{\footnotesize
\setlength{\tabcolsep}{4pt}
\setlength{\LTpre}{10pt}
\setlength{\LTpost}{6pt}
\begin{xltabular}{\textwidth}{@{}ll>{\raggedright\arraybackslash}X@{}}
\caption{Routing features used by each single-group adaptive router. The features in the ``All'' row of a router are used in every task family, and the features in the ``VQA'', ``Audio-QA'', and ``Video-QA'' rows are used only in that task family. The notation \texttt{a\_\{b, c\}} is short for the two features \texttt{a\_b} and \texttt{a\_c}, and $k \in \{0,1,2\}$ is the index of a region, segment, or clip. The feature names are the same as in Tables~\ref{tab:feature_definitions}, \ref{tab:audio_feature_definitions}, and~\ref{tab:video_feature_definitions}.}
\label{tab:simple-features} \\
\toprule
\textbf{Single-group Adaptive Router} & \textbf{Task Family} & \textbf{Features Used} \\
\midrule
\endfirsthead
\toprule
\textbf{Single-group Adaptive Router} & \textbf{Task Family} & \textbf{Features Used} \\
\midrule
\endhead
\midrule
\multicolumn{3}{r}{\itshape Continued on next page} \\
\endfoot
\bottomrule
\endlastfoot
Question-type routing & All      & \texttt{target\_nouns}, \texttt{target\_nouns\_len}, \texttt{target\_nouns\_\{0,1,2,3\}}, \texttt{demand}, \texttt{demand\_question\_category}, \texttt{demand\_question\_category\_len}, \texttt{demand\_question\_category\_\{0,1\}}, \texttt{demand\_target\_noun\_count}, \texttt{demand\_relation\_cue\_count} \\
& VQA      & \texttt{demand\_requires\_\{localization, interaction, spatial\_reasoning, color\_or\_texture, ocr, counting\}} \\
& Audio-QA & \texttt{demand\_requires\_\{speech\_content, sound\_identification, temporal\_reasoning, emotion\_or\_tone, speaker\_identification, counting\}} \\
& Video-QA & \texttt{demand\_requires\_\{action\_recognition, temporal\_reasoning, spatial\_reasoning, appearance\_recognition, ocr, counting\}} \\
\midrule
Missing-evidence-only routing & All      & \texttt{text\_evidence\_to\_confirm} \\
& VQA      & \texttt{image}, \texttt{image\_crops}, \texttt{image\_crop\_$k$}, \texttt{image\_crop\_$k$\_box}, \texttt{image\_crop\_$k$\_box\_\{0,1,2,3\}}, \texttt{image\_crop\_$k$\_\{confidence, caption, caption\_relevance, contains\_targets, matched\_nouns\}}, \texttt{image\_crop\_\{0,1\}\_matched\_nouns\_\{len, 0\}}, \texttt{image\_crop\_0\_matched\_nouns\_1}, \texttt{router\_best\_caption\_relevance}, \texttt{router\_crop\_relevance\_\{mean, std, min, max, distribution\}}, \texttt{router\_low\_relevance\_crop\_ratio}, \texttt{router\_crop\_confidence\_variance}, \texttt{router\_crop\_rel\_dist\_\{very\_low, low, medium, high\}} \\
& Audio-QA & \texttt{audio}, \texttt{audio\_clips}, \texttt{audio\_clip\_$k$}, \texttt{audio\_clip\_$k$\_\{start, end, confidence, description, description\_relevance, contains\_targets, matched\_nouns\}}, \texttt{audio\_clip\_\{0,1\}\_matched\_nouns\_\{len, 0\}}, \texttt{audio\_clip\_0\_matched\_nouns\_1}, \texttt{router\_best\_description\_relevance}, \texttt{router\_clip\_relevance\_\{mean, std, min, max, distribution\}}, \texttt{router\_low\_relevance\_clip\_ratio}, \texttt{router\_clip\_confidence\_variance}, \texttt{router\_clip\_rel\_dist\_\{very\_low, low, medium, high\}} \\
& Video-QA & \texttt{video}, \texttt{video\_clips}, \texttt{video\_clip\_$k$}, \texttt{video\_clip\_$k$\_\{start, end, confidence, description, description\_relevance, contains\_targets, matched\_nouns\}}, \texttt{video\_clip\_\{0,1\}\_matched\_nouns\_\{len, 0\}}, \texttt{video\_clip\_0\_matched\_nouns\_1}, \texttt{router\_best\_description\_relevance}, \texttt{router\_clip\_relevance\_\{mean, std, min, max, distribution\}}, \texttt{router\_low\_relevance\_clip\_ratio}, \texttt{router\_clip\_confidence\_variance}, \texttt{router\_clip\_rel\_dist\_\{very\_low, low, medium, high\}} \\
\midrule
Confidence-only routing & All & \texttt{text}, \texttt{text\_initial\_answer}, \texttt{text\_confidence}, \texttt{text\_reasoning}, \texttt{text\_signals}, \texttt{text\_signals\_\{has\_hedges, hedge\_terms\_found, cites\_visual\_evidence, generic\_priors, prior\_terms\_found, prior\_terms\_len, prior\_terms\_0\}}, \texttt{text\_self\_critique}, \texttt{text\_self\_critique\_neg\_reasoning}, \texttt{router\_text\_confidence}, \texttt{router\_confidence\_calibration\_flag} \\
\midrule
Cost-aware heuristic & All      & \texttt{router}, \texttt{router\_graph\_\{target\_node\_coverage, subgraph\_confidence, relation\_match, irrelevant\_node\_ratio, spurious\_relation\_risk, clutter\_risk, attribute\_loss\_risk, marginal\_relation\_gain\}}, \texttt{similarity\_q\_\{initial, critique, evidence, all\}} \\
& VQA      & \texttt{router\_image\_marginal\_coverage\_gain}, \texttt{router\_text\_image\_overlap}, \texttt{router\_image\_\{irrelevant\_entity\_risk, distracting\_context\_risk\}}, \texttt{router\_image\_graph\_overlap}, \texttt{similarity\_q\_caption}, \texttt{similarity\_image\_\{cropmean, cropmax\}} \\
& Audio-QA & \texttt{router\_audio\_marginal\_coverage\_gain}, \texttt{router\_text\_audio\_overlap}, \texttt{router\_audio\_\{irrelevant\_content\_risk, distracting\_content\_risk\}}, \texttt{router\_audio\_graph\_overlap}, \texttt{similarity\_q\_description} \\
& Video-QA & \texttt{router\_video\_marginal\_coverage\_gain}, \texttt{router\_text\_video\_overlap}, \texttt{router\_video\_\{irrelevant\_entity\_risk, distracting\_context\_risk\}}, \texttt{router\_video\_graph\_overlap}, \texttt{similarity\_q\_description}, \texttt{similarity\_video\_\{clipmean, clipmax\}} \\
\end{xltabular}
}

\subsection{Robustness to Incorrect Routing Features}
\label{subsec:robust-simple}

In this section, we compare the robustness of \name and the four single-group adaptive routers to incorrect routing features. For each single-groupr adaptive router, we flip, mask, or shift the feature group used by the single-group adaptive router and apply the same change to \name. We report the accuracy and token cost of each router before and after each change in Table~\ref{tab:robust-simple}, averaged over the eight benchmarks. 

To evaluate the robustness to incorrect question type features, we flip the question type features. We find that the accuracy of question-type routing drops by 1.1\%, whereas the accuracy of \name drops by only 0.3\%, since question-type routing relies on the question type features alone, whereas \name still uses the missing-evidence statement and the other three feature groups. To evaluate the robustness to incorrect confidence features, we shift the confidence features in the uncertainty cues by a Gaussian perturbation $\epsilon \sim \mathcal{N}(0, 0.1^2)$. We find that the accuracy of confidence-only routing drops by 0.2\% and the accuracy of \name drops by 0.1\%, but confidence-only routing is less accurate than \name even before the change, at 73.0\% compared with 74.8\%. 

To evaluate the robustness to incorrect missing-evidence features, we mask the missing-evidence features, which are the missing-evidence statement and the evidence availability cues. We find that the accuracy of missing-evidence-only routing drops by 1.7\% while its token cost decreases from 18,214 to 17,823, which indicates under-fusion. In contrast, the accuracy of \name drops by only 0.4\% while the token cost of \name increases from 16,247 to 17,491, since the accuracy-first objective of \name minimizes token cost only among the combinations expected to answer correctly, and \name therefore keeps accuracy by spending more tokens when the evidence availability cues become unreliable. 

To evaluate the robustness to incorrect additional modality gain cues, we shift the additional modality gain cues by the same Gaussian perturbation. We find that the accuracy of \name stays at 74.8\%, and only the token cost of \name increases from 16,247 to 17,003. The additional modality gain cues are used to drop a modality that adds little new evidence, and therefore \name drops fewer modalities and uses more tokens when these cues are shifted. However, the other three feature groups still decide which modalities the instance requires, so the accuracy of \name stays unchanged, whereas the accuracy of the cost-aware heuristic drops by 0.3\% from an already lower 72.3\%. 

Therefore, the two most accurate single-group adaptive routers, question-type routing and missing-evidence-only routing, are the least robust, and the other two single-group adaptive routers are more stable but less accurate. Under each of the four changes, the accuracy of \name drops less than the accuracy of the corresponding single-group adaptive router, since \name uses all four feature groups and does not rely on one feature group.

\begin{table}[ht]
\centering
\caption{Accuracy (Acc) and average token cost (Tkn) of each single-group adaptive router and of \name before $\rightarrow$ after the feature group used by the single-group adaptive router is flipped, masked, or shifted, averaged over the eight benchmarks. Accuracy is reported as a percentage (\%), and Tkn denotes the average per-instance token cost.}
\label{tab:robust-simple}
\vspace{0.5em}
\setlength{\tabcolsep}{5pt}
\begin{tabular}{llcc}
\toprule
\textbf{Changed Features} & \textbf{Method} & \textbf{Acc} & \textbf{Tkn} \\
\midrule
\multirow{2}{*}{Question type features}        & Question-type routing         & 73.7 $\rightarrow$ 72.6 & 17,784 $\rightarrow$ 18,766 \\
                                               & \name                         & 74.8 $\rightarrow$ 74.5 & 16,247 $\rightarrow$ 16,903 \\
\midrule
\multirow{2}{*}{Confidence features}           & Confidence-only routing       & 73.0 $\rightarrow$ 72.8 & 16,707 $\rightarrow$ 17,346 \\
                                               & \name                         & 74.8 $\rightarrow$ 74.7 & 16,247 $\rightarrow$ 16,377 \\
\midrule
\multirow{2}{*}{Missing-evidence features}     & Missing-evidence-only routing & 73.5 $\rightarrow$ 71.8 & 18,214 $\rightarrow$ 17,823 \\
                                               & \name                         & 74.8 $\rightarrow$ 74.4 & 16,247 $\rightarrow$ 17,491 \\
\midrule
\multirow{2}{*}{Additional modality gain cues} & Cost-aware heuristic          & 72.3 $\rightarrow$ 72.0 & 15,693 $\rightarrow$ 16,981 \\
                                               & \name                         & 74.8 $\rightarrow$ 74.8 & 16,247 $\rightarrow$ 17,003 \\
\bottomrule
\end{tabular}
\end{table}

\section{FaithScore Case Studies}
\label{sec:faithscore-cases}
In this section, we present three VQA cases that show why a FaithScore above a sufficient level is enough and why maximizing FaithScore is not desirable. FaithScore counts a claim as grounded only when the claim is directly observable in the input, so a correct inference drawn from the input lowers FaithScore in the same way as a hallucination. However, many questions ask for something that is not directly observable, such as the role of a person or the kind of a dish, and can only be answered by an inference from observed evidence. A correct answer to such a question therefore needs both observed evidence and an inference drawn from that evidence. The observed evidence keeps the answer from being pure speculation, and the inference is what the question asks for. As a result, a FaithScore above a sufficient level shows that the answer is anchored in observed evidence, whereas a higher FaithScore would remove the inference that the question requires.

\begin{figure}[!t]
\centering
\small
\begin{subfigure}{\textwidth}
\begin{minipage}[c]{0.48\textwidth}\centering\includegraphics[width=\linewidth,height=6cm,keepaspectratio]{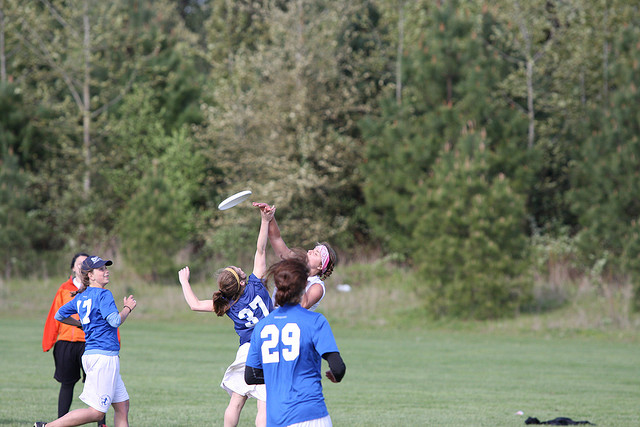}\end{minipage}\hfill
\begin{minipage}[c]{0.49\textwidth}
\textbf{Question:} What would be the reason a person on the field is dressed in orange and black?\\[2pt]
\textbf{Answer of \name:} referee\\[2pt]
\textbf{Observed evidence:} The person wears orange and black, and the players wear blue.\\[2pt]
\textbf{Penalized inference:} The person has a different role from the players and is therefore a referee.\\[2pt]
\textbf{FaithScore:} 0.70
\end{minipage}
\caption{OKVQA}
\end{subfigure}\\[10pt]
\begin{subfigure}{\textwidth}
\begin{minipage}[c]{0.48\textwidth}\centering\includegraphics[width=\linewidth,height=6cm,keepaspectratio]{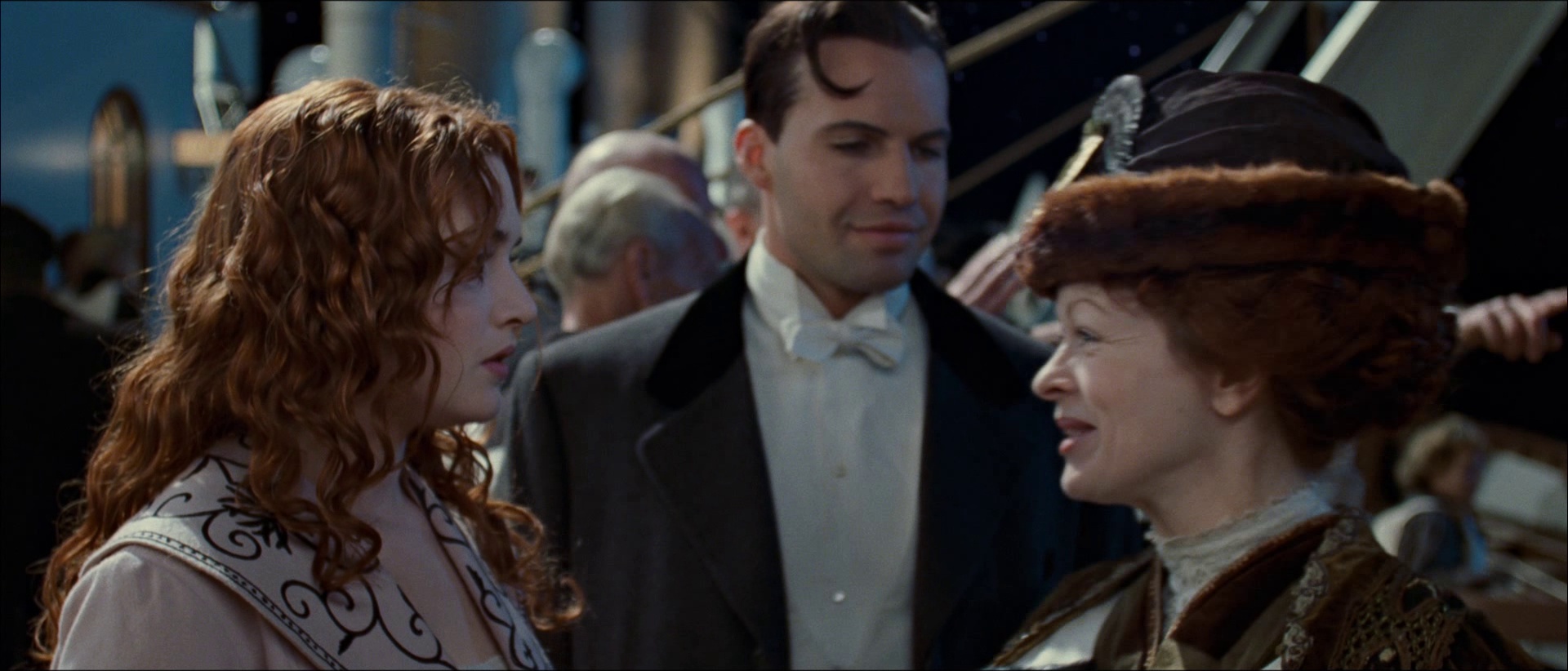}\end{minipage}\hfill
\begin{minipage}[c]{0.49\textwidth}
\textbf{Question:} Why does Riley look so smug?\\[2pt]
\textbf{Answer of \name:} Riley approves and agrees with what Jackie is saying.\\[2pt]
\textbf{Observed evidence:} Riley looks at Jackie with a slight smile.\\[2pt]
\textbf{Penalized inference:} The slight smile and direct gaze indicate approval.\\[2pt]
\textbf{FaithScore:} 0.69
\end{minipage}
\caption{VCR}
\end{subfigure}\\[10pt]
\begin{subfigure}{\textwidth}
\begin{minipage}[c]{0.48\textwidth}\centering\includegraphics[width=\linewidth,height=6cm,keepaspectratio]{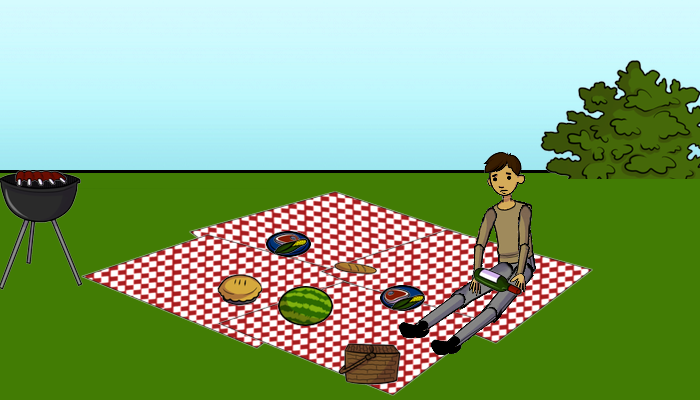}\end{minipage}\hfill
\begin{minipage}[c]{0.49\textwidth}
\textbf{Question:} What kind of pie is it?\\[2pt]
\textbf{Answer of \name:} apple\\[2pt]
\textbf{Observed evidence:} The pie is brown and has lines on top.\\[2pt]
\textbf{Penalized inference:} A brown pie with lines on top is an apple pie.\\[2pt]
\textbf{FaithScore:} 0.67
\end{minipage}
\caption{VQA-v2}
\end{subfigure}
\caption{Three VQA cases in which \name answers correctly but receives a low FaithScore. Each case shows the image, the question, the answer of \name, the evidence observed in the image, the inference that FaithScore penalizes, and the FaithScore.}
\label{fig:faithscore-cases}
\end{figure}

We show the image, the question, the answer of \name, the observed evidence, the penalized inference, and the FaithScore of each case in Fig.~\ref{fig:faithscore-cases}. In the first case from OKVQA, the question asks why a person on the field is dressed in orange and black. \name observes that the person wears orange and black whereas the players wear blue, infers that the person has a different role from the players, and answers ``referee''. In the second case from VCR, the question asks why Riley looks smug. \name observes that Riley looks at Jackie with a slight smile, infers approval from the smile and the gaze, and answers that Riley approves and agrees with what Jackie is saying. In the third case from VQA-v2, the question asks what kind of pie is shown. \name observes that the pie is brown and has lines on top, infers that a brown pie with lines on top is an apple pie, and answers ``apple''. In all three cases, the answer is correct, the observed evidence is visible in the image, and the inference is required, since the role of a person, the attitude of a person, and the kind of a pie are not directly observable. However, FaithScore counts each inference as an ungrounded claim, and the three cases receive FaithScores of only 0.70, 0.69, and 0.67. Therefore, a low FaithScore in these cases reflects an inference that the question requires rather than a hallucination. This result also explains why \name obtains higher accuracy than $\mathrm{T+X}$ on VQA-v2 with a slightly lower FaithScore in Section~\ref{sec:faithscore}.

\end{document}